\documentclass[a4paper,11pt]{article}
\pdfoutput=1

\usepackage{array}
\usepackage{soul}
\usepackage{xcolor}
\usepackage{mathtools}
\usepackage{amsmath}
\usepackage{amsfonts}
\usepackage{amssymb}
\usepackage{graphicx,bm}
\usepackage{a4wide}
\usepackage{booktabs}
\usepackage[small,bf]{caption}
\usepackage{cite}
\usepackage[colorlinks]{hyperref}
\usepackage{ulem,xpatch}

\allowdisplaybreaks
\numberwithin{equation}{section}

\DeclareMathOperator{\ci}{\text{i}}

\DeclareMathOperator{\diff}{\text{d}}
\DeclareMathOperator\erf{erf}

\newcommand\ddfrac[2]{\frac{\displaystyle #1}{\displaystyle #2}}

\newcommand{\rsun}{r_\text{sun}}

\newcolumntype{d}[1]{D{.}{\cdot}{#1} }

\begin{document}

\begin{titlepage}

\vspace*{-15mm}
\begin{flushright}
NCTS-HEPAP/2201
\end{flushright}
\vspace*{0.7cm}

\begin{center}
{
\bf\LARGE 
Taking Neutrino Pictures via Electrons
}
\\[8mm]
Guey-Lin Lin$^{\, a,}$ \footnote{E-mail: \texttt{glin@nycu.edu.tw}},
Thi Thuy Linh Nguyen$^{\, a,}$ \footnote{E-mail: \texttt{thuylinh.sc09@nycu.edu.tw}},
Martin Spinrath$^{\, b, c}$ \footnote{E-mail: \texttt{spinrath@phys.nthu.edu.tw}},
Thi Dieu Hien Van$^{\, a,}$ \footnote{E-mail: \texttt{hieniop96.sc09@nycu.edu.tw}} 
\\and 
Tse-Chun Wang$^{\, d,}$ \footnote{E-mail: \texttt{tsechunwang@mx.nthu.edu.tw}}
\\[1mm]
\end{center}

\vspace*{0.50cm}
\centerline{$^{a}$ \it Institute of Physics, National Yang Ming Chiao Tung University, Hsinchu 300, Taiwan}
\vspace*{0.2cm}
\centerline{$^{b}$ \it Department of Physics, National Tsing Hua University, Hsinchu, 30013, Taiwan}
\vspace*{0.2cm}
\centerline{$^{c}$ \it Center for Theory and Computation,
National Tsing Hua University, Hsinchu 300, Taiwan}
\vspace*{0.2cm}
\centerline{$^{d}$ \it Physics Division, National Center for Theoretical Sciences, Taipei 10617, Taiwan}
\vspace*{1.20cm}

\begin{abstract}
\noindent
In this paper we discuss the prospects to take
a picture of an extended neutrino source, i.e., 
resolving its angular neutrino luminosity distribution.
This is challenging since neutrino directions
cannot be directly measured but only estimated
from the directions of charged particles they interact with in the detector
material. This leads to an intrinsic blurring effect.
We first discuss the problem
in general terms and then apply our insights to solar
neutrinos scattering elastically with electrons. Despite the aforementioned
blurring we show how with high
statistics and precision the original neutrino distributions
could be reconstructed.
\end{abstract}

\end{titlepage}
\setcounter{footnote}{0}

\section{Introduction}
\label{sec:Introduction}

Astronomy is one of the oldest scientific endeavours of mankind
which received a huge boost in the last century with the invention of telescopes
sensitive to photons over a huge range of energies far above and below
the optical spectrum which is also still improving and expanding. This progress is
based to a large extent on an improved physical understanding of the properties
and the nature of \textit{light} or \textit{photons} and its interactions with
matter.
This advancement allowed to resolve the true nature of planets, moons and other objects
in the solar system which seem point-like without telescopes.

But photons are not the only messengers of astronomical sources
routinely being studied. In this paper we will focus on neutrinos
which similarly have gained importance for astrophysics due to the
improved and still constantly improving understanding of their nature
and origin driven by particle physics.
Latest since the observation of solar neutrinos by the Homestake experiment - see, e.g., \cite{Cleveland:1998nv} for a review on
the experiment and \cite{Gann:2021ndb} for a recent review of solar neutrinos  -
and
the detection of neutrinos from supernova SN1987A \cite{Kamiokande-II:1987idp,Bionta:1987qt,Alekseev:1988gp}
they have established themselves as an important tool in
astrophysics. Also the studied energy range of neutrinos has expanded
significantly culminating in neutrino telescopes such as IceCube~\cite{IceCube:2011ucd} and KM3NET~\cite{KM3Net:2016zxf}.
The simultaneous observation of photons and neutrinos from the
same astrophysical sources is the template for modern multimessenger 
astroparticle physics
which recently gained an additional boost by the
observation of
gravitational waves.

Neutrinos are interesting as astrophysical messengers since they travel
almost freely not taking part in electromagnetic and strong
interactions, for a recent review, see, for instance,~\cite{GalloRosso:2018omb}.
Indeed, very often neutrinos are produced inside of stars and then escape
with small distortions compared to other messengers. It takes extreme events like
supernovae to affect neutrinos substantially on their path and even
then neutrinos are the
first messengers escaping and giving valuable information on the
earliest stages of a supernova~\cite{Burrows:2012ew,Nakamura:2016kkl,Muller:2019upo}. For that reason they are being used,
for instance, in
an alarm system to point other telescopes to interesting astrophysical events~\cite{SNEWS:2020tbu,LIGOScientific:2017ync,Beacom:1998fj,Ando:2001zi,Mukhopadhyay:2020ubs,Chen:2021hkl}.

The drawback is that neutrinos 
are also hard to detect for the very same
reasons. And in all experiments
their detection happens indirectly via elastic or inelastic
processes sometimes eradicating substantial information about the original
neutrino energy, direction and flavour~\cite{SNEWS:2020tbu,Beacom:1998fj,Ando:2001zi,Davis:2016hil,Chen:2021hkl}. In this paper we will focus on elastic
scattering processes which still introduce quite some uncertainty in the
reconstruction of the incident neutrino energy and direction in an
individual event
and we compare our results to the results from \cite{Davis:2016hil}.
Since there are new
experiments coming up or proposed with improved
detection capabilities and analysis techniques
(\textit{e.g.}, the Hyper-Kamiokande (HyperK) experiment~\cite{Hyper-Kamiokande:2018ofw}, the Jiangmen Underground Neutrino Observatory (JUNO)~\cite{JUNO:2015sjr,JUNO:2020hqc}, the SNO$+$ experiment~\cite{SNO:2021xpa}, IceCube-GEN2~\cite{IceCube:2014gqr}, the Precision IceCube Next Generation Upgrade (PINGU)~\cite{IceCube-PINGU:2014okk}, KM3NeT~\cite{KM3Net:2016zxf}, etc), we think it is the right time
to go beyond point-like neutrino sources in the sky and study
if we can resolve their structures.

In this work we establish
a framework which would allow, in principle,
to reconstruct
the picture of an extended source in ``neutrino light'' and
apply this framework explicitly to solar neutrinos which are understood
rather well from theory and experiment. As a guiding question we will ask
if it is possible to distinguish a point source and a ring-shaped source
the size of the sun for neutrinos with an energy spectrum and flavour
composition as solar neutrinos.
We will focus here on the electron picture and not
explicitly reconstruct the neutrino picture since, in particular,
the inclusion of experimental
uncertainties make a reconstruction of the solar neutrino pictures with
a sufficient resolution quite difficult at this point.
We find that the distinction between a point and a ring source
just from the electron picture is rather hard but
theoretically possible.

Nevertheless, if we would be able to reconstruct the neutrino picture
successfully with high precision -- and we consider our work presented
here as a step in this direction -- it would offer
a way to distinguish different solar models and help to reconstruct a
full three-dimensional model of the sun including things like temperature
and pressure profiles.
They are connected to the
radial neutrino flux profiles of the sun which we will translate
into two-dimensional angular luminosity distributions
of neutrinos, the neutrino pictures.
Different from Ref.~\cite{Davis:2016hil}, for a better understanding of the
physics inside the sun (or other sources), we analyse both angular and energy distributions
of solar neutrino events, in terms of the
double differential event rate later in this work. The double differential event
rate ideally helps us as well to measure the spatial separation of different neutrino sources.
From our results it will also be clear  for which kind of sources it could
lead to a less blurry picture than for solar neutrinos.

This paper is arranged as follows:
We will first derive in Sec.~\ref{sec:Framework} the theoretical framework
and show how the
observed energy and angular distributions of scattered electrons
are related to the energy and angular distribution of the incoming neutrino
flux. Interestingly, they are related by an invertible integral transformation
which involves the differential cross section as well.
In Sec.~\ref{sec:Simple} we apply this framework to some simple
toy examples and also make a first step towards answering, how large the
neutrino picture of the sun is. In Sec.~\ref{sec:SunInNeutrinos} we discuss
the fully realistic case for $^8$B, hep and pep solar neutrinos and
their pictures. Then we illustrate how those pictures
could be affected by experimental errors in
Sec.~\ref{sec:ExperimentalErrors} before we summarise and conclude
in Sec.~\ref{sec:Summary}. We also provide extensive appendices
with additional details on the derivation of the
formulas in the main text.

\section{Theoretical Framework}
\label{sec:Framework}

To take a neutrino picture of any extended neutrino source
one has to consider the scattering of neutrinos
in the detector material. Here we will focus on solar neutrinos
with an energy around 10~MeV and adopt neutrino electron elastic
scattering  $\nu+e^-\rightarrow \nu + e^-$.
All flavors of neutrinos (and for other sources anti-neutrinos)
can be detected by this interaction, though the cross sections
depend on flavour and chirality. We will show here how to reconstruct
the angular and energy distribution of neutrinos by
measuring the direction and energy of the scattered electrons.
In this section, we first build the theoretical framework for this
methodology and then study applications in the rest of the paper.

A very similar exercise was done for directional Dark Matter (DM)
detection and we will follow this approach  as presented in \cite{Gondolo:2002np}
closely here.
There is one major difference between DM and our case that is that the incoming neutrinos here
are highly relativistic and hence practically massless while the incoming
DM particles there are assumed to be non-relativistic.

\begin{figure}
\centering
\includegraphics[width=0.45\linewidth]{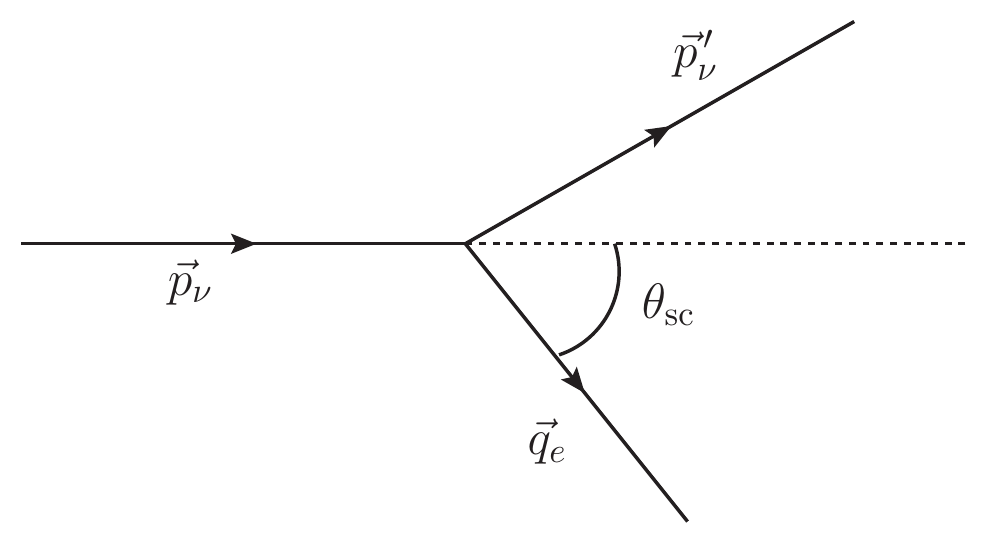}    
\caption{Sketch to  clarify the kinematics used in Sec.~\ref{sec:Framework}.
\label{fig:Kinematics}}
\end{figure}

We want to assume that the electrons in our target are at rest and free (we
neglect binding energies here). The relativistic energy momentum conservation
then reads, cf.~Fig.~\ref{fig:Kinematics},
\begin{align}
\sqrt{m_\nu^2 + p_\nu^2} + m_e &= \sqrt{m_\nu^2 +  {p'}_\nu^2} + \sqrt{m_e^2 + q_e^2} \;, \\
p'_\nu \cos \theta_{\text{sc}} &=  p_\nu - q_e \cos \theta_{\text{sc}} \;, \\
p'_\nu \sin \theta_{\text{sc}} &= q_e \sin \theta_{\text{sc}} \;,
\end{align}
where $m_\nu$, $p_\nu$ and $p'_\nu$ are neutrino mass and modulus of three-momentum
before and after scattering. The mass of the electron is $m_e$ and the modulus of its three-momentum
after scattering is $q_e$. The scattering angle with respect to the incoming neutrino
direction is defined as  $\theta_{\text{sc}}$.

We can sum the squares of the last two equations to get
\begin{align}
{p'}_\nu^2 = (p_\nu - q_e \cos \theta_{\text{sc}})^2 + (q_e \sin \theta_{\text{sc}})^2 = p_\nu^2 - 2 p_\nu q_e \cos \theta_{\text{sc}} + q_e^2
\end{align}
and use this equation to eliminate ${p'}_\nu^2$ from the first equation.
Using the relativistic approximation $m_\nu \ll p_\nu \approx E_\nu$
we then find
\begin{align}
\cos \theta_{\text{sc}} &=
\frac{(E_e - m_e)(E_\nu + m_e) }{E_\nu \, \sqrt{E_e^2 - m_e^2}}
= \hat{p}_\nu \cdot \hat{q}_e  \;, \label{eq:costhscEe}
\end{align}
where we have noted that the cosine of the scattering angle between the incoming neutrino
direction and the outgoing electron direction can be written
in a coordinate independent way as the scalar product of the normalised momentum vectors
of the incoming neutrino and the outgoing electron.
The outgoing electron energy is restricted to the range
\begin{align}
m_e \leq E_e \leq \frac{2 E_\nu^2 + 2E_\nu \, m_e + m_e^2}{2 E_\nu + m_e} \equiv E_e^{\text{max}} \label{eq:EeMax}
\end{align}
corresponding to $0 \leq \cos \theta_{\text{sc}} \leq 1 \;$.
We want to remind here that $E_e$ is the total electron energy after
scattering including the electron mass, and the electron
kinetic energy $T_e = E_e - m_e$ as labeled.

Our aim here is to derive an expression for
the double differential event rate, in events per unit time
per unit detector mass, differentiated with respect to the electron energy
$E_e$ and electron recoil
direction $\hat{q}_e$
\begin{equation}
\frac{\diff R}{\diff E_e \diff \Omega_e} \;,
\end{equation}
where $\diff \Omega_e$ denotes the infinitesimal solid angle around
the direction $\hat{q}_e$.

The double differential rate follows from the double differential cross section
\begin{equation}
\frac{\diff \sigma}{\diff E_e \diff \Omega_e} \;.
\end{equation}
In the scattering processes we consider here the electron energy $E_e$ and
scattering direction are not independent from each other. But we can impose
their relation explicitly with a $\delta$-function using  eq.~\eqref{eq:costhscEe} 
\begin{equation}
\frac{\diff \sigma}{\diff E_e \diff \Omega_e} = \frac{\diff \sigma}{\diff E_e} \frac{1}{2 \pi} \delta\left(\hat{p}_\nu \cdot \hat{q}_e - \frac{(E_e - m_e)(E_\nu + m_e) }{E_\nu \, \sqrt{E_e^2 - m_e^2}}  \right) \;.
\end{equation} 
This is crucial and was already done in
\cite{Gondolo:2002np} which nevertheless looks
somewhat different there due to the different kinematics.

Now to go from the double differential cross section to the double
differential event rate we have to multiply the first with the number
of electrons in the detector $N_e$, divide by its mass $M_D$ and multiply
with the flux of incoming neutrinos.
Again, since we talk about incoming relativistic particles
we will deviate from the presentation in \cite{Gondolo:2002np}
somewhat.
The number flux element in our case is given by
\begin{equation}
\frac{\diff F(E_\nu,\Omega_\nu)}{\diff E_\nu \diff \Omega_\nu} \diff E_\nu \diff \Omega_\nu \;.
\end{equation}

For solar neutrinos the distributions of energy and direction of
incoming neutrinos are to a good approximation
independent from each other, and we can write
\begin{equation}
\frac{\diff F(E_\nu,\Omega_\nu)}{\diff E_\nu \diff \Omega_\nu} \diff E_\nu \diff \Omega_\nu =
f_0 \frac{\diff \epsilon(E_\nu)}{\diff E_\nu} \frac{ \diff \lambda(\Omega_\nu)}{\diff \Omega_\nu} \diff E_\nu \diff \Omega_\nu  \;,
\end{equation}
where the energy and angular distributions itself are normalised, i.e.,
\begin{equation}
\int \diff E_\nu \frac{\diff \epsilon(E_\nu)}{\diff E_\nu} = 1 \quad \text{ and } \quad
\int \diff \Omega_\nu \frac{ \diff \lambda(\Omega_\nu)}{\diff \Omega_\nu} = 1\;.
\end{equation}

Then
\begin{align}
\frac{\diff R}{\diff E_e \diff \Omega_e} &= \frac{N_e}{M_D} \int  \diff E_\nu \diff \Omega_\nu \frac{\diff \sigma(E_e, E_\nu)}{\diff E_e \diff \Omega_e} \frac{\diff F(E_\nu,\Omega_\nu)}{\diff E_\nu \diff \Omega_\nu}  \nonumber\\
&= \frac{N_e \, f_0}{2 \pi \,M_D} \int \diff E_\nu \, \frac{\diff \epsilon(E_\nu)}{\diff E_\nu}  \frac{\diff \sigma (E_e, E_\nu)}{\diff E_e}   \,  \nonumber\\
&\quad \times \int \diff \Omega_\nu \frac{ \diff \lambda(\Omega_\nu)}{\diff \Omega_\nu} \delta\left( \hat{p}_\nu \cdot \hat{q}_e - \frac{(E_e - m_e)(E_\nu + m_e) }{E_\nu \, \sqrt{E_e^2 - m_e^2}}  \right) \;.
\label{eq:KeyFormula}
\end{align}
This equation is the key formula which we will use throughout the rest of
the paper.

What is interesting about this is that by measuring the double differential
scattered electron rate, it is possible to infer the original
neutrino distributions. For DM the author of \cite{Gondolo:2002np}
has shown that the double differential event rate is proportional to the Radon
transform of the DM velocity distribution.
In our case this is again not quite that easy but we can 
find an analogous result.

We start from rewriting our key formula
eq.~\eqref{eq:KeyFormula}
and after using  that $E_\nu = |\vec{p}_\nu|$ and $E_\nu^2 \diff E_\nu \diff \Omega_\nu = \diff^3 \vec{p}_\nu$,
\begin{align}
\frac{\diff R}{\diff E_e \diff \Omega_e}  
&= \frac{N_e \, f_0}{2 \pi \,M_D} \int \diff^3 \vec{p}_\nu \frac{\diff \epsilon}{\diff |\vec{p}_\nu|}  \frac{ \diff \lambda(\Omega_\nu)}{\diff \Omega_\nu}  \frac{1}{|\vec{p}_\nu|} \frac{\diff \sigma}{\diff E_e}(E_e, |\vec{p}_\nu|)   \,  \nonumber\\
&\quad \times  \delta\left(  \vec{p}_\nu \cdot \hat{q}_e - \frac{(E_e - m_e)(|\vec{p}_\nu| + m_e)}{\sqrt{E_e^2 - m_e^2}}  \right) \;.
\end{align}
Now we can define the generalised distribution analogous to the DM velocity distribution
\begin{equation}
g(E_e,\vec{p}_\nu) \equiv f_0  \frac{\diff \epsilon}{\diff |\vec{p}_\nu|}  \frac{ \diff \lambda(\Omega_\nu)}{\diff \Omega_\nu}  \frac{1}{|\vec{p}_\nu|} \frac{\diff \sigma}{\diff E_e}(E_e, |\vec{p}_\nu|)
\end{equation}
and the vector
\begin{equation}
\vec{w} = \frac{ (E_e - m_e)(E_\nu + m_e)}{\sqrt{E_e^2 - m_e^2} } \hat{q}_e \;,
\end{equation}
which is parallel to the electron recoil direction $\hat{q}_e$ but has a different length, $w$.
We arrive at the double differential rate
\begin{align}
\frac{\diff R}{\diff E_e \diff \Omega_e}  
&= \frac{N_e }{2 \pi \,M_D} \int \diff^3 \vec{p}_\nu \, g(E_e,\vec{p}_\nu) \, \delta\left(  \vec{p}_\nu \cdot \hat{w} - w \right) = \frac{N_e }{2 \pi \,M_D} \widehat{g}(w,\hat{w}) \;.
\end{align}
At this point the Radon transform of the distribution function
\begin{equation}
\widehat{g}(w,\hat{w}) =  \int \diff^3 \vec{p}_\nu \, g(E_e,\vec{p}_\nu) \, \delta\left(  \vec{p}_\nu \cdot \hat{w} - w \right)
\end{equation}
appears. In the case of DM the distribution function $g$ was actually just
the DM velocity distribution which does not depend on the electron energy.
In our case, this is more complicated due to the relativistic nature of
the incoming neutrinos and since we also do not assume the cross section to be
approximately constant over the considered momentum range as can be done
for many DM models.

In \cite{Gondolo:2002np} the author also mentions some ways to invert
the Radon transform. This is indeed a common, well studied problem
since this transform plays an important role, for instance, in
medical imaging. Using that inversion techniques one could hence
derive the function $g(E_e,\vec{p}_\nu)$ after measuring 
the double differential event rate. To get the angular distribution
of the neutrinos one would then still need to have knowledge of the
neutrino energy spectrum and the differential neutrino electron cross
section.
Monochromatic sources where all neutrinos have the same energy
would hence be particularly well suited subjects of study.

The cross section is well known and we use here the result from \cite{Marciano:2003eq}
which we have collected for the convenience
of the reader in App.~\ref{app:CrossSections} rewritten in terms of $E_e$
as we will need it.

\section{Some simplified examples}
\label{sec:Simple}

Before we consider the fully realistic example of solar neutrinos we want to
discuss two simple toy examples first. The ultimate goal of this exercise
would be to resolve a non-trivial angular distribution of a given neutrino
source. To do this we want to study two extreme examples, a point and a
ring source (we study radial symmetric sources throughout this paper).
The question is then, could we actually distinguish a point from a
ring source?

\subsection{Monochromatic Point vs.\ Ring Source}
\label{sec:PointvsRing}

Let us first discuss the point source. In the examples we will
always choose the lab frame such that the
center of the radial symmetric source is in positive $z$-direction (at $\cos \theta_{e,\nu} =1$).
We also assume for now that the
neutrino source is monochromatic, that means all neutrinos have
the same energy, $E_\nu = E_0$.

For the point source we then assume the following angular and energy distributions
\begin{equation}
\frac{ \diff \lambda_p(\Omega_\nu)}{\diff \Omega_\nu} = \frac{1}{2 \pi } \delta(\cos \theta_\nu - 1) 
\text{ and } \frac{\diff \epsilon_p(E_\nu)}{\diff E_\nu} = \delta(E_\nu - E_0) \;.
\end{equation}
We can then derive the double differential rate as (we provide more details
of the derivations of the relevant formulas in this section in App.~\ref{app:RingPointApp})
\begin{align}
\frac{\diff R_p}{\diff E_e \diff \Omega_e} &= 
\frac{N_e \, f_0}{2 \pi \,M_D}  \frac{\diff \sigma}{\diff E_e} (E_e, E_\nu = E_0)  \,  \delta\left(  \cos \theta_e - \frac{(E_e - m_e)(E_0 + m_e)}{E_0 \sqrt{E_e^2 - m_e^2}}  \right) \nonumber\\
&\equiv \gamma_p \, \delta\left(  \cos \theta_e - \frac{(E_e - m_e)(E_0 + m_e)}{E_0 \sqrt{E_e^2 - m_e^2}}  \right) \;,
\label{eq:gammap}
\end{align}
where we have defined $\gamma_p$ which we will use later.
At this point one could choose to integrate over the electron energy or
the electron angular distribution depending on what one is interested in.
To show an explicit example, we integrate over the electron energy
from a threshold value $E_e^{\text{thr}}$ to infinity
to get the angular distribution
\begin{align}
\frac{\diff R_p}{\diff \Omega_e} &= \int_{E_e^{\text{thr}}}^\infty \diff E_e \frac{\diff R_p}{\diff E_e \diff \Omega_e} \nonumber\\
&= \frac{N_e \, f_0}{2 \pi \,M_D}  \frac{\diff \sigma}{\diff E_e} \left(E_\nu = E_0,  E_e = E_e^{(p)}  \right) \, \theta(E_e^{(p)} - E_e^{\text{thr}}) \frac{E_0 (E_e^{(p)} + m_e) \sqrt{(E_e^{(p)})^2 - m_e^2} }{ m_e (E_0 + m_e) }  \;,
\end{align}
where
\begin{equation}
E_e^{(p)} = m_e \frac{ E_0^2 ( \cos^2 \theta_e + 1)  + 2 \, m_e \, E_0 + m_e^2 }{ E_0^2 \sin^2 \theta_e + 2 \, m_e \, E_0 + m_e^2} \;.
\end{equation}
To go from this solid angle distribution to the simple angular distribution
$\diff R_p / \diff \cos \theta_e$ one would simple have to multiply
by $2 \pi$ due to the radial symmetry.

Let us now consider a monochromatic ring source, i.e., 
\begin{equation}
\frac{ \diff \lambda_r(\Omega_\nu)}{\diff \Omega_\nu} = \frac{1}{2 \pi } \delta\left(\cos \theta_\nu - c_r \right) 
\text{ and } \frac{\diff \epsilon_r(E_\nu)}{\diff E_\nu} = \delta(E_\nu - E_0) \;,
\end{equation}
where we do not yet specify the opening angle of the ring but assume it is
small, $1 \gg 1- c_r^2 > 0$.
We then find for the double differential event rate
\begin{align}
\frac{\diff R_r}{\diff E_e \diff \Omega_e} &= 
\frac{N_e \, f_0}{4 \pi^2 \,M_D}   
\int_0^{2 \pi} \diff \phi_\nu \,  \frac{\diff \sigma}{\diff E_e} (E_e,E_\nu = E_0) \delta\left( (\hat{p}_\nu \cdot \hat{q}_e)_r  - \frac{(E_e - m_e)(E_0 + m_e)}{E_0 \sqrt{E_e^2 - m_e^2}}  \right)  \;.
\label{eq:RingdRdOmegae}
\end{align}
where
\begin{equation}
(\hat{p}_\nu \cdot \hat{q}_e)_r = \sqrt{1-c_r^2}  \cos \phi_\nu \cos \phi_e \sin \theta_e + \sqrt{1-c_r^2} \sin \phi_\nu \sin \phi_e \sin \theta_e  + c_r \cos \theta_e \;, 
\end{equation}
and for the angular distribution
\begin{align}
\frac{\diff R_r}{\diff \Omega_e} &= \int_{E_e^{\text{thr}}}^\infty \diff E_e \frac{\diff R}{\diff E_e \diff \Omega_e} \nonumber\\
&= \frac{N_e \, f_0}{4 \pi^2 \,M_D}  \int_0^{2 \pi} \diff \phi_\nu \,   \frac{\diff \sigma}{\diff E_e} (E_e = E_e^{(r)}, E_\nu 
= E_0 )  \,  \theta(E_e^{(r)} - E_e^{\text{thr}}) \nonumber\\
&\quad \times \frac{E_0 (E_e^{(r)} + m_e) \sqrt{(E_e^{(r)})^2 - m_e^2} }{ m_e (E_0 + m_e) }\;,
\end{align}
where
\begin{equation}
\label{eq:hatERing}
E_e^{(r)} = m_e \frac{ E_0^2 ( (\hat{p}_\nu \cdot \hat{q}_e)_r^2 + 1)  + 2 \, m_e \, E_0 + m_e^2 }{E_0^2 ( 1- (\hat{p}_\nu \cdot \hat{q}_e)_r^2) + 2 \, m_e \, E_0 + m_e^2} \;.
\end{equation}
So here we will still have to integrate over $\phi_\nu$ which is not trivial
since $E_e^{(r)}$ depends still on $\phi_\nu$ via $(\hat{p}_\nu \cdot \hat{q}_e)_r$.

At this point we can already make some comparison of a point with a ring source
to get a better understanding of the prospects to distinguish them.
We consider a ring source with an opening angle as wide as
the optical disc of the sun. That is
\begin{equation}
\theta_{\text{sun}} \equiv \arccos \sqrt{ 1 - \frac{\rsun^2 }{D_\text{sun}^2} } \approx 4.65 \times 10^{-3} \approx 0.267^\circ \;,
\end{equation}
where we have used as radius of the sun $\rsun = 696,340$~km and
the distance to the sun is on average $D_\text{sun} = 149,600,000$~km.
Note that solar neutrinos are produced inside of the sun in a
sphere with a radius smaller than $\rsun$. We will come back to
this later.

\begin{figure}
\centering
\includegraphics[width=0.7\linewidth]{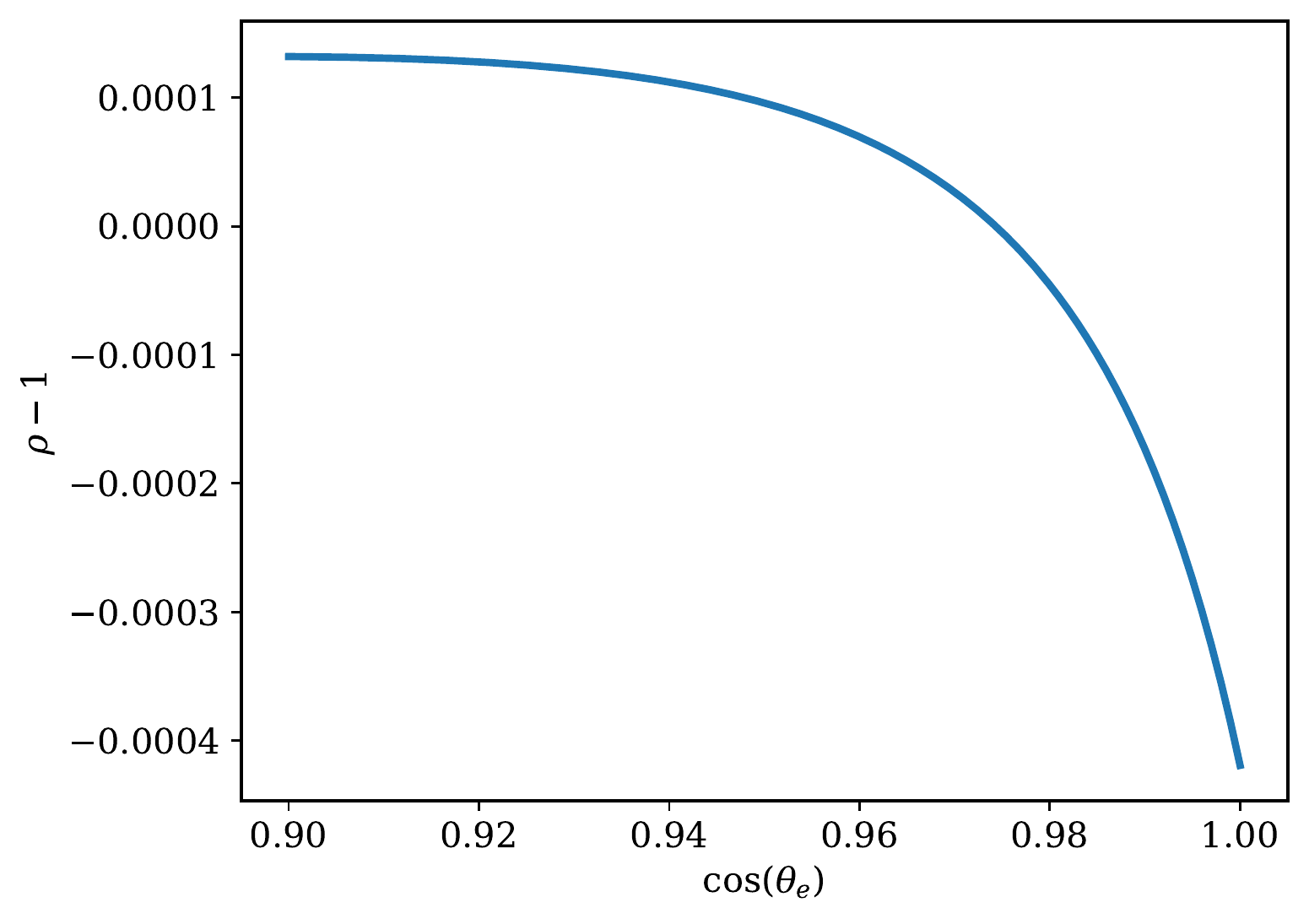}
\caption{
Ratio of the differential electron scattering rate for a
ring and a point source. Assuming a fixed neutrino energy
of 10~MeV and an opening angle of the ring of $\theta_{\text{sun}}$. The differences are at the sub-permil level.
\label{fig:PointvsRing_10_Rsun}}
\end{figure}

In Fig.~\ref{fig:PointvsRing_10_Rsun} we show the ratio
\begin{equation}
    \rho = \ddfrac{\diff R_r/\diff \Omega_e}{ \diff R_p/\diff \Omega_e} \;.
\end{equation}
That is the ratio of the differential rates for a ring and a point
source. We used a fixed neutrino energy of 10~MeV and the opening
angle of the ring is equivalent to the opening angle of the sun in
the sky.
For the cross section we also just consider electron neutrino electron scattering
at this point. 
We see that it is not easy to distinguish the ring and the point source
in this simplified setup.
Their difference being only at the sub-permil level. That is actually
not surprising considering the rather tiny opening angle of the ring
compared to the typical scattering angle. The electron
picture blurs such small neutrino pictures quite
drastically at this energy.

\begin{figure}
\centering
\includegraphics[width=0.7\linewidth]{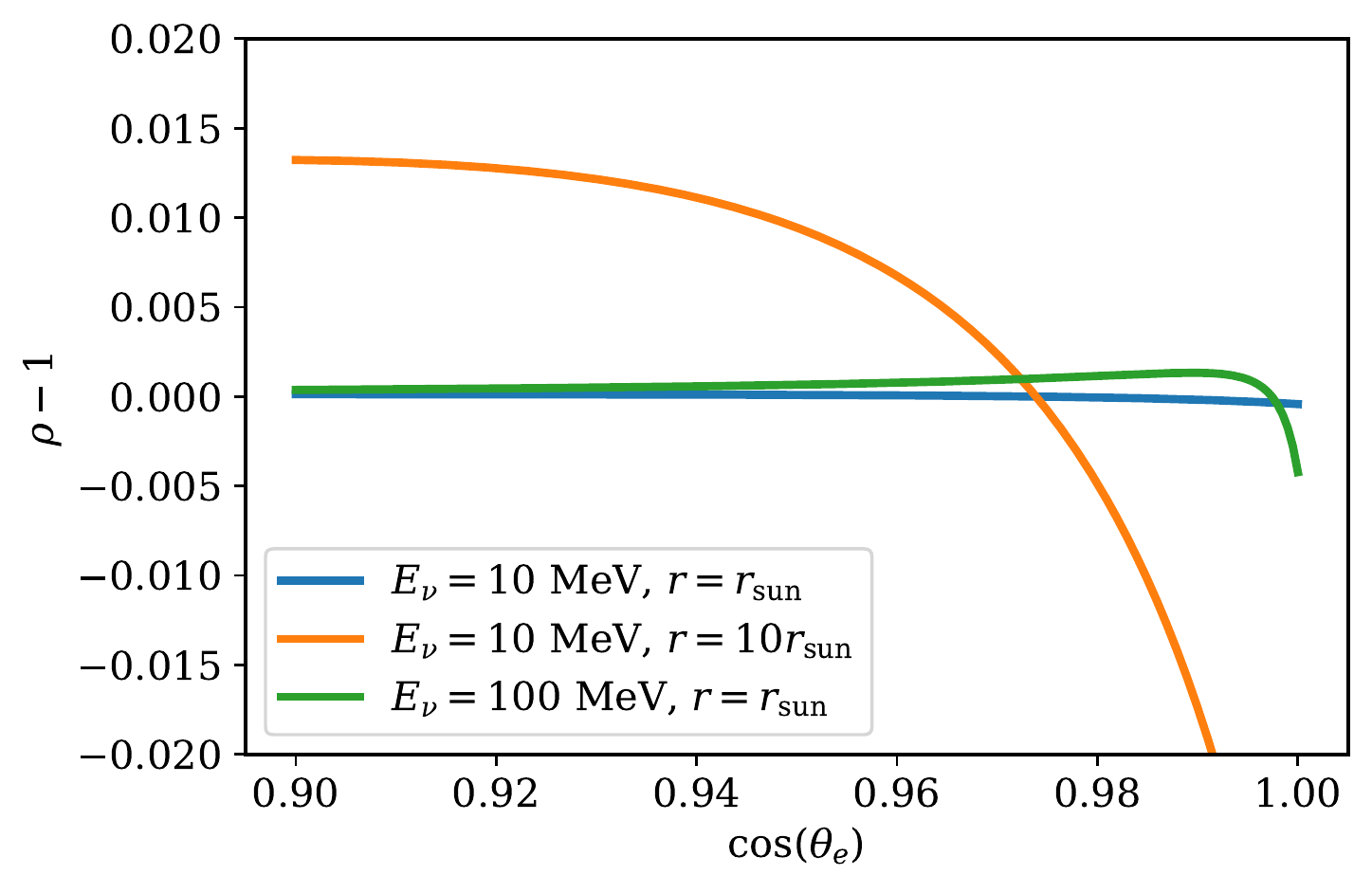}
\caption{
Ratio of the differential electron scattering rate for a
ring and a point source. Assuming a fixed neutrino energy
of 100~MeV and a ring with the radius of the sun and 10~MeV on the right with a ten times larger radius. The differences are at the percent level.
\label{fig:PointvsRing_others}}
\end{figure}

For comparison we also show the same ratio but change some of the
parameters in Fig.~\ref{fig:PointvsRing_others}. We first
increase the neutrino energy to 100~MeV.
The higher the neutrino energy the stronger the correlation between
the electron and neutrino direction. Therefore the relative
deviation grows in particular for small scattering angles
and reaches the permil level.

Then we also increased the opening angle assuming
a ring-like object which is ten times larger than the sun
but again setting the neutrino energy to 10~MeV.
We can see even deviations at the percent level also distributed
over wider scattering angles which would make an experimental
distinction between a point and a ring source significantly easier.

\begin{figure}
\centering
\includegraphics[width=0.8\linewidth]{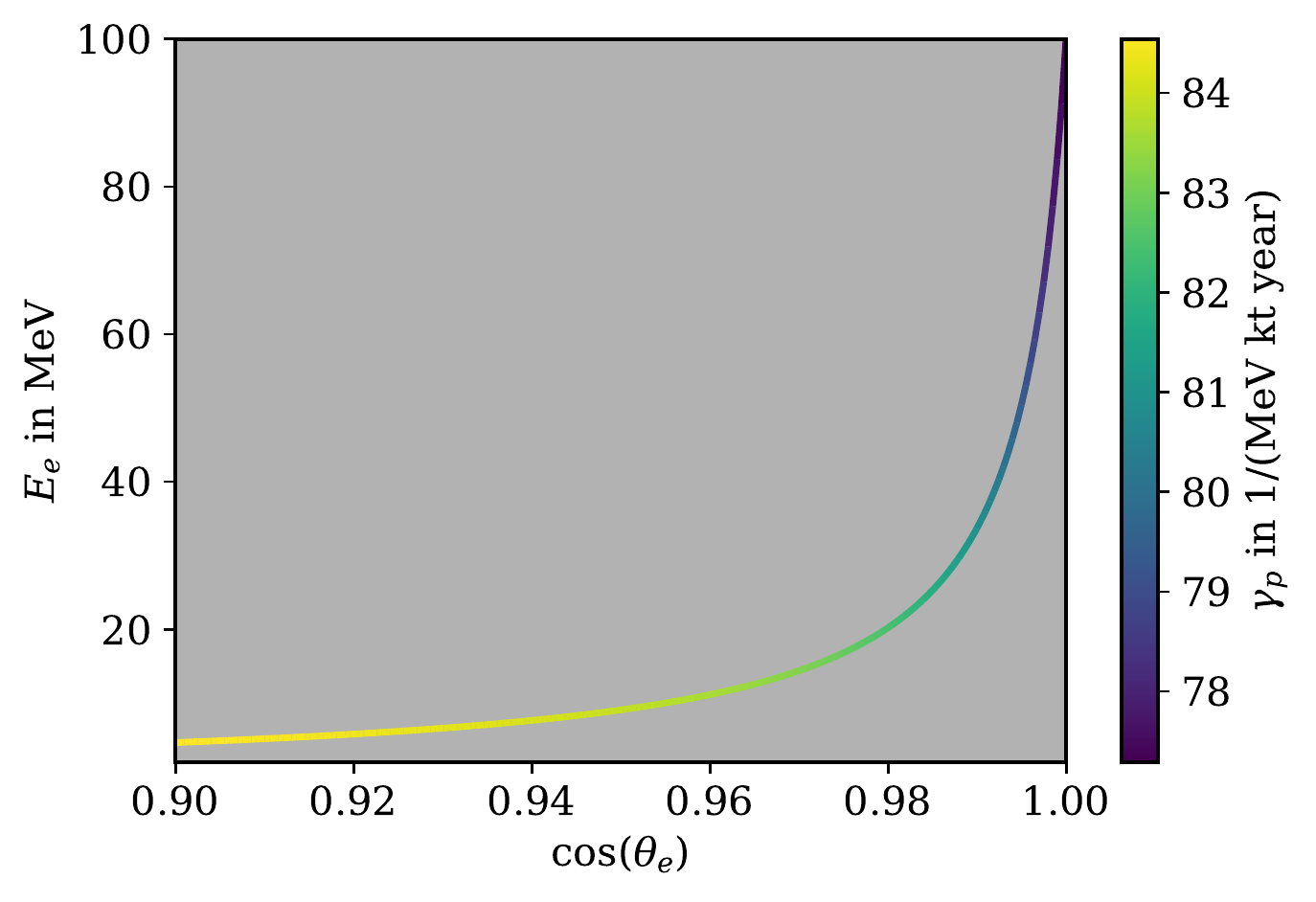}
\caption{
Representation of the double differential electron neutrino electron
scattering rate in the $\cos\theta_e$-$E_e$
plane for a point source with a neutrino energy of 100~MeV
and a neutrino flux of $5 \times 10^{6}$~cm$^{-2}$~s$^{-1}$.
\label{fig:Point_dRdEedOmegae}}
\end{figure}

Nevertheless, we want to propose here not to look at the angular or energy
rate distributions but at
the double differential rate instead.

For a monochromatic point source this is not quite trivial since
it is proportional to a $\delta$-function. Therefore we show
in Fig.~\ref{fig:Point_dRdEedOmegae}
the correlation between electron energy and scattering angle
for a point source for a fixed neutrino energy of 100~MeV
and the coefficient in front of the $\delta$-function, $\gamma_p$, c.f.~eq.~\eqref{eq:gammap}.
In the grey areas there are no events for a monochromatic point source.
We have assumed water as target material and a neutrino flux
of $f_0 = 5 \times 10^{6}$~cm$^{-2}$~s$^{-1}$. While the flux corresponds
roughly to the solar $^8$B neutrino flux we have increased the neutrino
energy to values larger than typical solar neutrinos to show
more clearly the difference to a ring source in that same plane.

\begin{figure}
\centering
\includegraphics[width=0.9\linewidth]{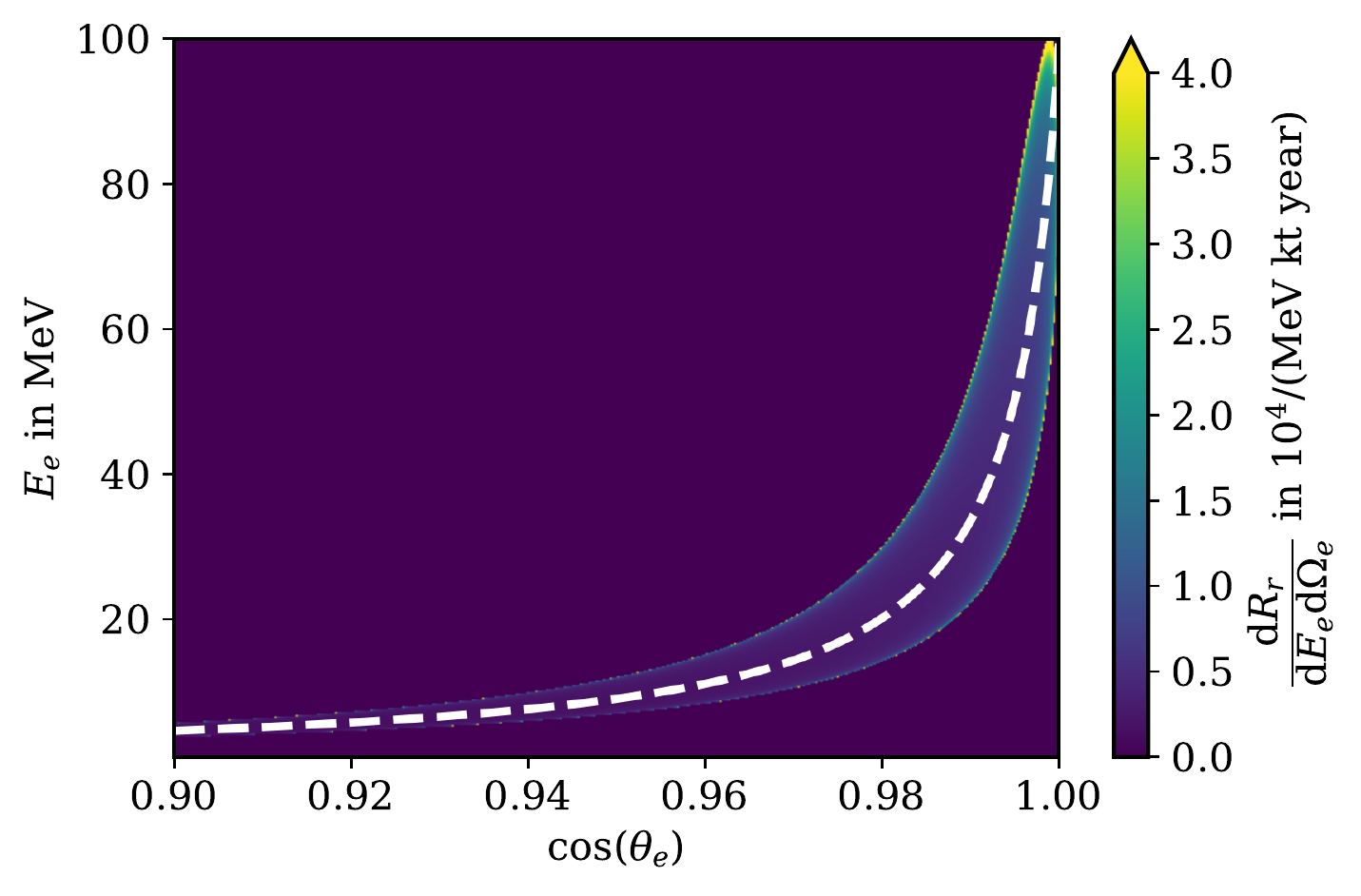} 
\caption{
Double differential scattering rate for a
ring source. We assume a fixed neutrino energy
of 100~MeV, a neutrino flux of $5 \times 10^{6}$~cm$^{-2}$~s$^{-1}$
and the ring is ten times larger than the sun for
better visibility. For comparison we show
as a white dashed line the correlation between electron
energy and scattering angle for a point source,
cf.~Fig.~\ref{fig:Point_dRdEedOmegae}.
To increase the visibility of the extremely thin distribution
we cap the color scale at $4 \times 10^4$ 1/(MeV~kt~year).
\label{fig:Ring_dRdEedOmegae}}
\end{figure}

We show that difference in Fig.~\ref{fig:Ring_dRdEedOmegae}
where we plot the double differential rate for a ring source
ten times the size of the sun and a neutrino energy of
100~MeV for water as target material. For a comparison
we show the line for a 100~MeV point source. We can 
see that the ring in the two-dimensional plane looks
clearly different from a point source. Interestingly
the event rates get larger towards the outside of the
distribution which might be possible to be exploited
experimentally.

Nevertheless, this have all been toy examples so far.
In a realistic source the geometry is usually not simple
and also very often the neutrino energy is not fixed
to a concrete value. 

\subsection{Towards Realistic Pictures}

Before going to the full pictures we want to study one intermediate step first.
We will again compare a ring and a point source, but use the energy distribution, total flux
and flavour information as expected for $^8$B neutrinos which are most well
studied in water Cherenkov detectors.

Starting again from our key formula, eq.~\eqref{eq:KeyFormula},
we can find the double differential rate for a point source with $^8$B energy
spectrum
\begin{align}
\frac{\diff R^p_{\text{8B}}}{\diff E_e \diff \Omega_e} 
&= \frac{N_e \, f_0^{\text{8B}}}{2 \pi \,M_D} \frac{\diff \epsilon_{\text{8B}}}{\diff E_\nu}(E_\nu^p)  \frac{\diff \sigma (E_e, E_\nu^p)}{\diff E_e}
 m_e \left| \frac{(E_e - m_e) \sqrt{E_e^2 - m_e^2}}{(E_e - m_e - \cos \theta_e \sqrt{E_e^2 - m_e^2})^2} \right|\;,
\end{align}
where
\begin{equation}
    E_\nu^p = \frac{m_e (E_e - m_e)}{m_e - E_e + \cos \theta_e \sqrt{E_e^2 - m_e^2}} \;.
\end{equation}
For a ring source with $^8$B energy
spectrum we find
\begin{align}
\frac{\diff R^r_{\text{8B}}}{\diff E_e \diff \Omega_e} &= 
\frac{N_e \, f_0^{\text{8B}}}{4 \pi^2 \,M_D} \int_0^{2\pi} \diff \phi_\nu \frac{\diff \epsilon_{\text{8B}}}{\diff E_\nu}(E_\nu^r)  \frac{\diff \sigma (E_e, E_\nu^r)}{\diff E_e}
\nonumber\\
&\phantom{\frac{N_e \, f_0^{\text{8B}}}{4 \pi^2 \,M_D} \int_0^{2\pi} \diff \phi_\nu \frac{\diff \epsilon_{\text{8B}}}{\diff E_\nu}(E_\nu^r) }
\times
 m_e \left| \frac{(E_e - m_e) \sqrt{E_e^2 - m_e^2}}{(E_e - m_e - (\hat{p}_\nu \cdot \hat{q}_e)_r \sqrt{E_e^2 - m_e^2})^2} \right| \;,
\end{align}
where
\begin{align}
    E_\nu^r &= \frac{m_e (E_e - m_e)}{m_e - E_e + (\hat{p}_\nu \cdot \hat{q}_e)_r \sqrt{E_e^2 - m_e^2}} \;,\\
    (\hat{p}_\nu \cdot \hat{q}_e)_r &= \sqrt{1-c_r^2}  \cos \phi_\nu \cos \phi_e \sin \theta_e + \sqrt{1-c_r^2} \sin \phi_\nu \sin \phi_e \sin \theta_e  + c_r \cos \theta_e \;.
\end{align}
More details on how to derive these formulas are given in App.~\ref{app:RingPointApp}.
For the energy distribution we take in both cases the GS98 prediction
taken from \cite{Vitagliano:2019yzm} with
$f_0^{\text{8B}} = 5.46 \times 10^6$~cm$^{-2}$~s$^{-1}$.

We will also include neutrino oscillations into the picture at this point.
As long as we consider elastic neutrino electron scattering $\mu$-
and $\tau$-neutrinos are indistinguishable, but the cross section
is different for them compared to electron neutrinos. To deal with this
we will replace
the differential cross section in the above formulas with
\begin{equation}
\frac{\diff \sigma (E_e, E_\nu)}{\diff E_e} = P_{ee}(E_\nu) \frac{\diff \sigma (\nu_e e \to \nu_e e)}{\diff E_e} + (1 - P_{ee}(E_\nu)) \frac{\diff \sigma (\nu_l e \to \nu_l e)}{\diff E_e} \;,
\end{equation}
where we have introduced the electron neutrino survival probability
for electron neutrinos produced in the sun,
which is generally energy dependent.
For simplicity, we will nevertheless
use the constant value
$P_{ee} = 0.37$~\cite{BOREXINO:2018ohr}
as an approximation since the energy dependence is not very strong
in the considered range.

\begin{figure}
\centering
\includegraphics[width=0.49\linewidth]{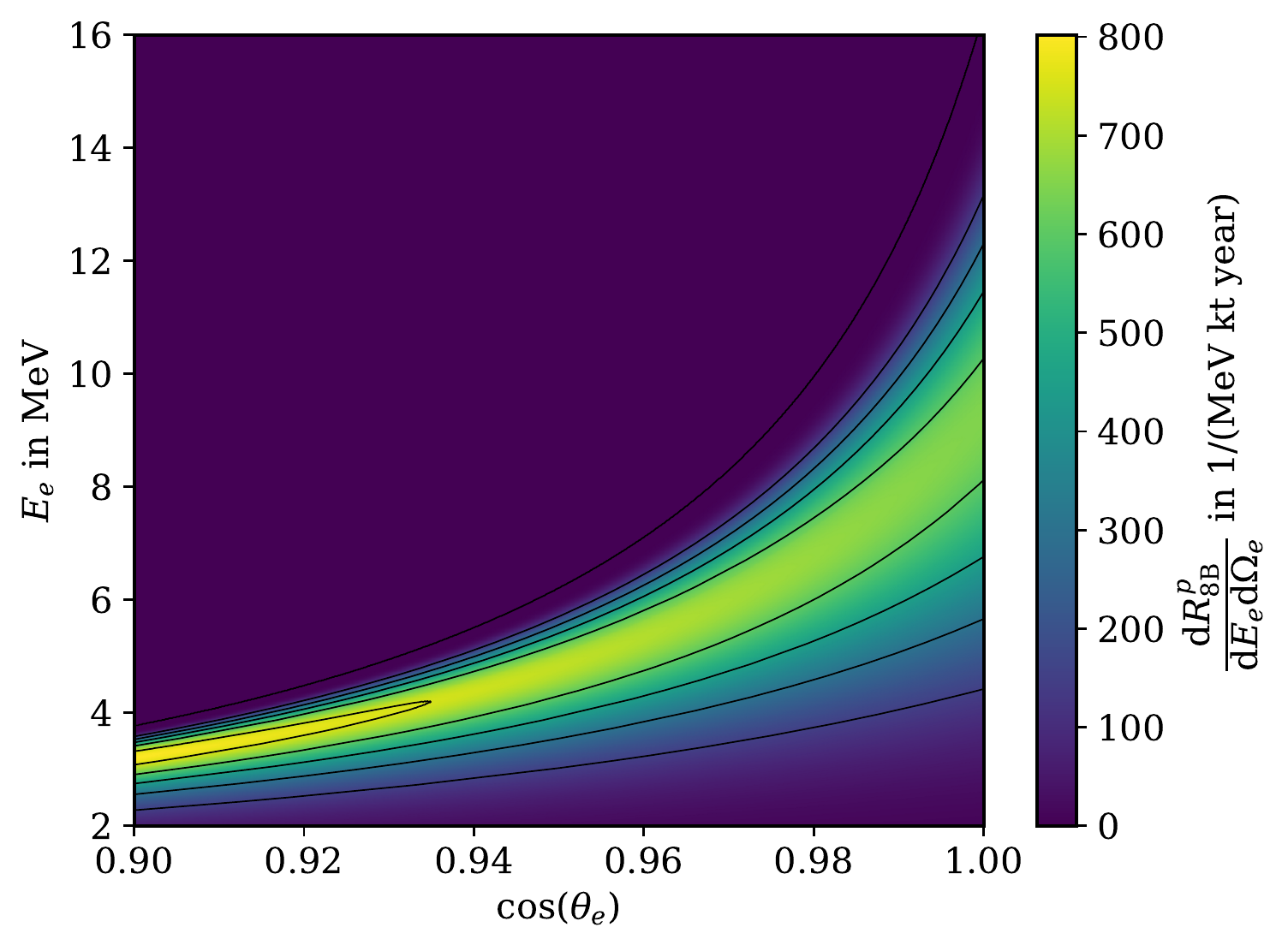} 
\includegraphics[width=0.49\linewidth]{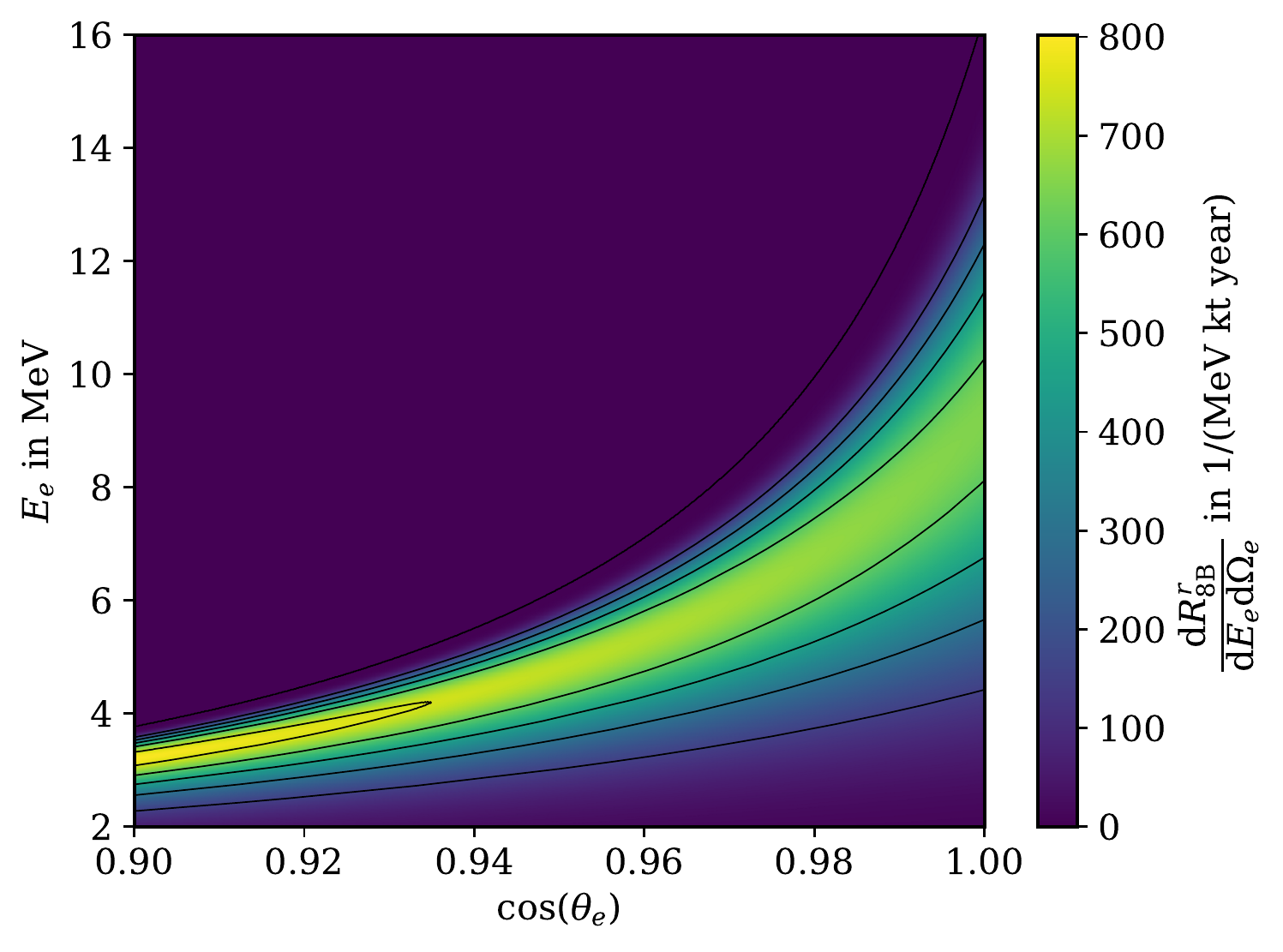}\\
\includegraphics[width=0.7\linewidth]{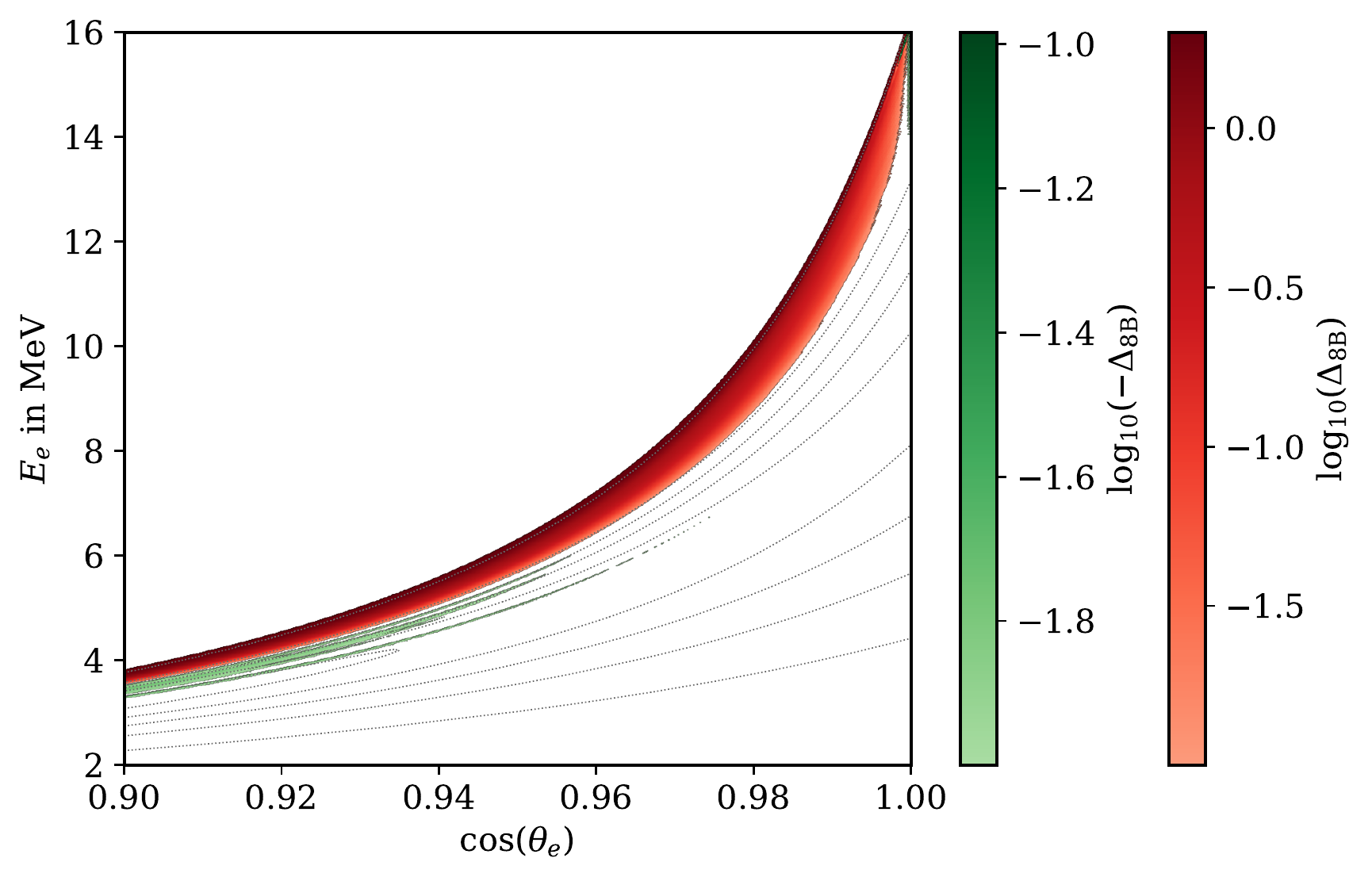}
\caption{
On top we show the double differential rate for neutrinos with energy spectrum
as the solar $^8$B neutrinos have. On the top left we assume the neutrinos
are produced in a point at the center of the sun and on the top right
we assume they are produced in a ring the size of the sun. On the bottom we show
the relative difference $\Delta_{\text{8B}}$ for the double differential event rates
between a point source and a ring source the size of the sun using otherwise identical
parameters. In the white regions the absolute, relative difference is either less than 1~\%
or not well-defined. In the red shaded regions the ring source
would lead to more events while in the green regions the point
source would result in more events.
For easier comparison we show the contour lines
of the point source in the comparison plot as dotted lines.
\label{fig:8B_Comparison}}
\end{figure}

In Fig.~\ref{fig:8B_Comparison} we first show the double differential distribution
in this setup for the point and the ring source on top.
Although we are considering a point
source here on the top left the double differential distribution is not
a line at all which is
due to the continuous energy distribution. The case for a ring the size of the
sun is extremely similar to the point source and hard to distinguish by eye.

We therefore defined the ratio
\begin{equation}
    \Delta_{\text{8B}} =  \ddfrac{ \frac{\diff R^r_{\text{8B}}}{\diff E_e \diff \Omega_e} - \frac{\diff R^p_{\text{8B}}}{\diff E_e \diff \Omega_e}}{ \frac{1}{2} \left( \frac{\diff R^r_{\text{8B}}}{\diff E_e \diff \Omega_e} + \frac{\diff R^p_{\text{8B}}}{\diff E_e \diff \Omega_e} \right) } \;,
\end{equation}
which is the difference of the two cases normalised by their average.
The result for $\log_{10} |\Delta_{\text{8b}}|$ is also shown
in Fig.~\ref{fig:8B_Comparison}.
In the white regions $|\Delta_{\text{8b}}|$ 
is either less than 1~\% or not well-defined.
The biggest difference is actually at the edge of the distributions,
which can be easily explained. 
For a ring source we expect to still see events for larger scattering angles
at the same electron energy since a larger source will result in a larger picture.
But the region where this difference is significant is unfortunately quite small
and also where the absolute event rate is not that large.

\section{Neutrino Pictures of the Sun via Electrons}
\label{sec:SunInNeutrinos}

So far we have discussed only point and ring sources.
Realistic solar models predict neither.
In this section we will discuss the fully realistic
case for $^8$B, hep and pep neutrinos.
These three neutrino sources are
dominantly produced in different regions of the sun.
The $^8$B neutrinos are being 
produced more towards the center of the sun while
hep and pep neutrinos have the peak production zone more outwards.
We choose these three sources for our benchmark study, not only because
they are well separated in the sun, but also due to the fact that they have different
features in energy. The hep neutrinos can reach up to $19$~MeV which is the highest
one among the three, and are expected to be identified by
HyperK~\cite{Hyper-Kamiokande:2018ofw}. However, for pep neutrinos
the energy is constant at $1.445$~MeV. By taking proper energy cuts, we can
separate them from each other.

\subsection{Angular Distributions from Radial Production Zones}
\label{sec:LuminostyProfile}

We begin the discussion with deriving the normalised neutrino
luminosity profile of the sun
from radial production zones. There have been many works studying
solar models and in our calculations we will use the results
from \cite{Vinyoles:2016djt}. For the considered neutrino fluxes
the results between the various calculations are usually similar.

What they usually provide in the literature are tables of the amount of neutrinos
produced in a given radial shell, i.e., they provide a
discrete version of a function $\diff j/\diff \zeta$,
where $\zeta = r/\rsun$. That implies $\diff j/\diff r = \rsun \diff j/\diff \zeta$.
In our
codes we import these data sets, interpolate and normalise them
such that
\begin{equation}
\int_0^1 \frac{\diff j}{\diff \zeta} \diff \zeta = \int_0^{\rsun} \frac{\diff j}{\diff r} \diff r = 1 \;.
\end{equation}
Note that here it is implicitly assumed
that the production zones are
spherically symmetric,
i.e., they only depend on the radius.
That also implies that the normalised neutrino production rate per unit volume at
any given point is given by
\begin{equation}
 J(\vec{r}) = \frac{1}{4\, \pi \, r^2} \frac{\diff j}{\diff r} \;.
\end{equation}

On earth we can of course just observe a two-dimensional
projection of the neutrino production distribution, which
we can calculate using the line of sight integral
\begin{align}
\frac{ \diff \lambda(\theta_{\nu},\phi_{\nu}) }{\diff \Omega_\nu}  = \mathcal{N}_\lambda \int \diff l  \; J(\vec{l}) \;,
\label{eq:Line_of_Sight}
\end{align}
where $l$ parametrises the line of sight and for convenience we introduce the
normalisation constant $\mathcal{N}_\lambda$ such that
\begin{equation}
\int \frac{ \diff \lambda }{\diff \Omega_\nu} \diff \Omega_\nu = 1 \;.
\end{equation}
Here we assumed that neutrinos just travel undisturbed
through the sun neglecting any attenuation and scattering effects
in the sun for simplicity and we have already mentioned
how we treat neutrino oscillations.

The observed total neutrino flux on earth is then given by
\begin{equation}
F = f_0 \int \diff E_\nu \int \diff \Omega_\nu \frac{\diff \epsilon}{\diff E_\nu}  \frac{ \diff \lambda }{\diff \Omega_\nu} 
\end{equation}
Note that we integrate here the absolute value of the flux over the source.
We do not take
the scalar product of the flux with an arbitrary detector area first and then integrate over the source.
That would lead to different values for the flux, especially for large, extended sources. For the sun, the
difference is not that large and we want to focus here on distributions and ignore
this subtle normalisation issue. We always specify what value
of $f_0$ we choose when needed.

\begin{figure}
\centering
\includegraphics[width=0.7\linewidth]{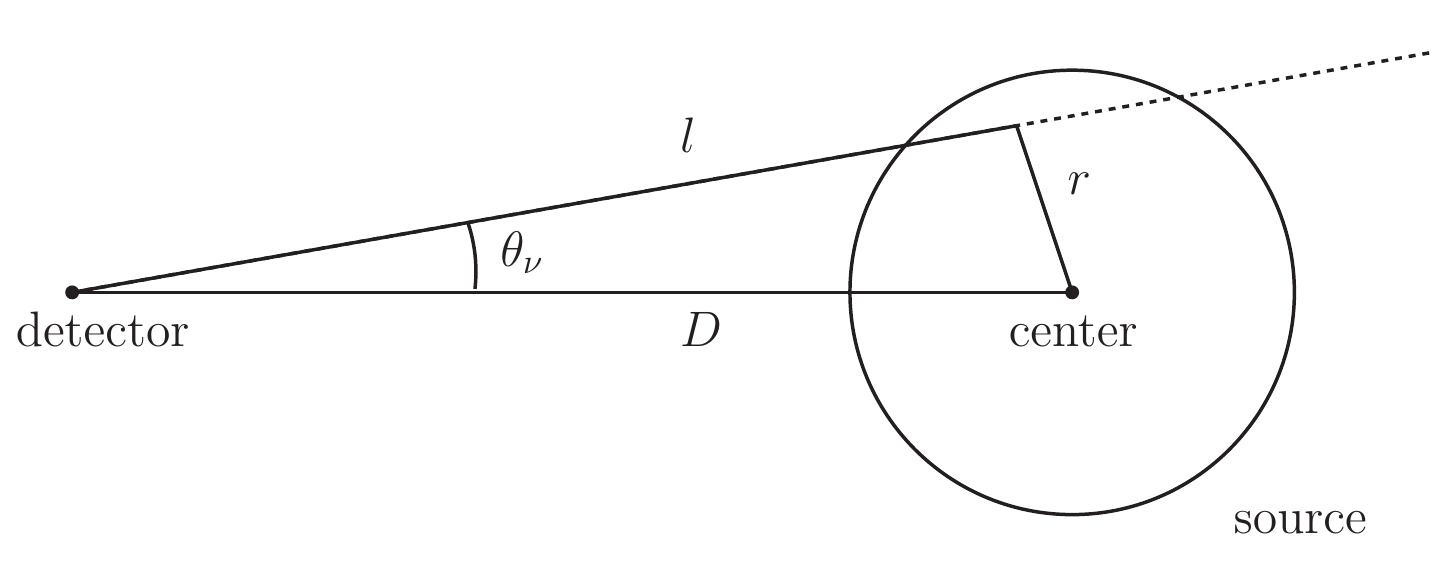} 
\caption{
We assume a radial symmetric source and a point-like detector.
Taking the line of sight integral we get the two-dimensional
luminosity profile of the source.
\label{fig:Line_of_Sight}}
\end{figure}

The problem in the derivation of the angular distribution
is that we cannot directly use eq.~\eqref{eq:Line_of_Sight}
since we do not have the function $J(\vec{l})$ or $\diff j(l)/\diff l$ readily
available. We only have the luminosity profile as a function 
of the normalised distance from the centre of the sun, 
i.e., we have $\diff j(\zeta)/\diff \zeta$ or $\rsun \diff j(r)/\diff r$.

From basic geometry, cf.~Fig.~\ref{fig:Line_of_Sight}, we know that
\begin{align}
r^2 &= D^2 + l^2 - 2 \, D \, l \cos \theta_\nu \;, \\
\Rightarrow r \diff r &= (l - D \cos \theta_\nu) \diff l \;,
\end{align}
where $D$ is the distance between the observer and the center
of the sun.
For future reference, we also note here that the maximum
observation angle
for a finite, radial symmetric source where the source center is
at $\theta_\nu = 0$ is given by
\begin{align}
\cos \theta_{\nu}^{\text{max}} =  \sqrt{1 - \frac{\rsun^2}{D^2}} \;.
\end{align}

We can now use the two solutions for $l(r)$
\begin{equation}
l(r) = \begin{cases}
l_1(r) = D \cos \theta_\nu - \sqrt{r^2 + \frac{D^2}{2} ( \cos(2\theta_\nu) -1 )  } \text{ for } l \leq D \cos \theta_\nu \;,\\
l_2(r) = D \cos \theta_\nu + \sqrt{r^2 + \frac{D^2}{2} ( \cos(2\theta_\nu) -1 )  } \text{ for } l > D \cos \theta_\nu
\;. 
\end{cases}
\end{equation}
The minimal and maximal line of sight are here
\begin{align}
 l_\text{max/min} \equiv D \cos \theta_\nu \pm \sqrt{\rsun^2 + \frac{D^2}{2} ( \cos(2\theta_\nu) -1 )  } \;.
\end{align}
We can then write
\begin{align}
 \frac{ \diff \lambda(\theta_{\nu},\phi_{\nu}) }{\diff \Omega_\nu}  &= \mathcal{N}_\lambda \int_{l_\text{min}}^{l_\text{max}} \diff l  \; J(\vec{l}) \nonumber\\
 &=  \mathcal{N}_\lambda \int_{l_\text{min}}^{D \cos \theta_\nu} \diff l_1  \; J(\vec{l}_1) 
   + \mathcal{N}_\lambda \int_{D \cos \theta_\nu}^{l_\text{max}} \diff l_2  \; J(\vec{l}_2) \nonumber\\
 &=   \mathcal{N}_\lambda \int_{\rsun}^{D \sin \theta_\nu} \frac{r \, \diff r}{l_1(r) - D \cos \theta_\nu}  \; \frac{1}{4 \, \pi \, r^2} \frac{\diff j(r)}{\diff r} \nonumber\\
 &\phantom{=}  + \mathcal{N}_\lambda \int_{D \sin \theta_\nu}^{\rsun} \frac{r \, \diff r}{l_2(r) - D \cos \theta_\nu}  \;  \frac{1}{4 \, \pi \, r^2} \frac{\diff j(r)}{\diff r} \nonumber\\
 &= \frac{\mathcal{N}_\lambda}{2 \, \pi} \int_{D \sin \theta_\nu}^{\rsun} \frac{\diff r}{r \sqrt{r^2 - D^2 \sin^2 \theta_\nu}} \frac{\diff j(r)}{\diff r}  \nonumber\\
 &= \frac{\mathcal{N}_\lambda}{2 \, \pi \, \rsun^2} \int^{1}_{D\sin \theta_\nu/\rsun} \frac{\diff \zeta}{\zeta \sqrt{\zeta^2 - (D/\rsun)^2 \sin^2 \theta_\nu}} \frac{\diff j(\zeta)}{\diff \zeta}  \;.
\end{align}

\begin{figure}
	\centering
	\includegraphics[width=0.8\textwidth]{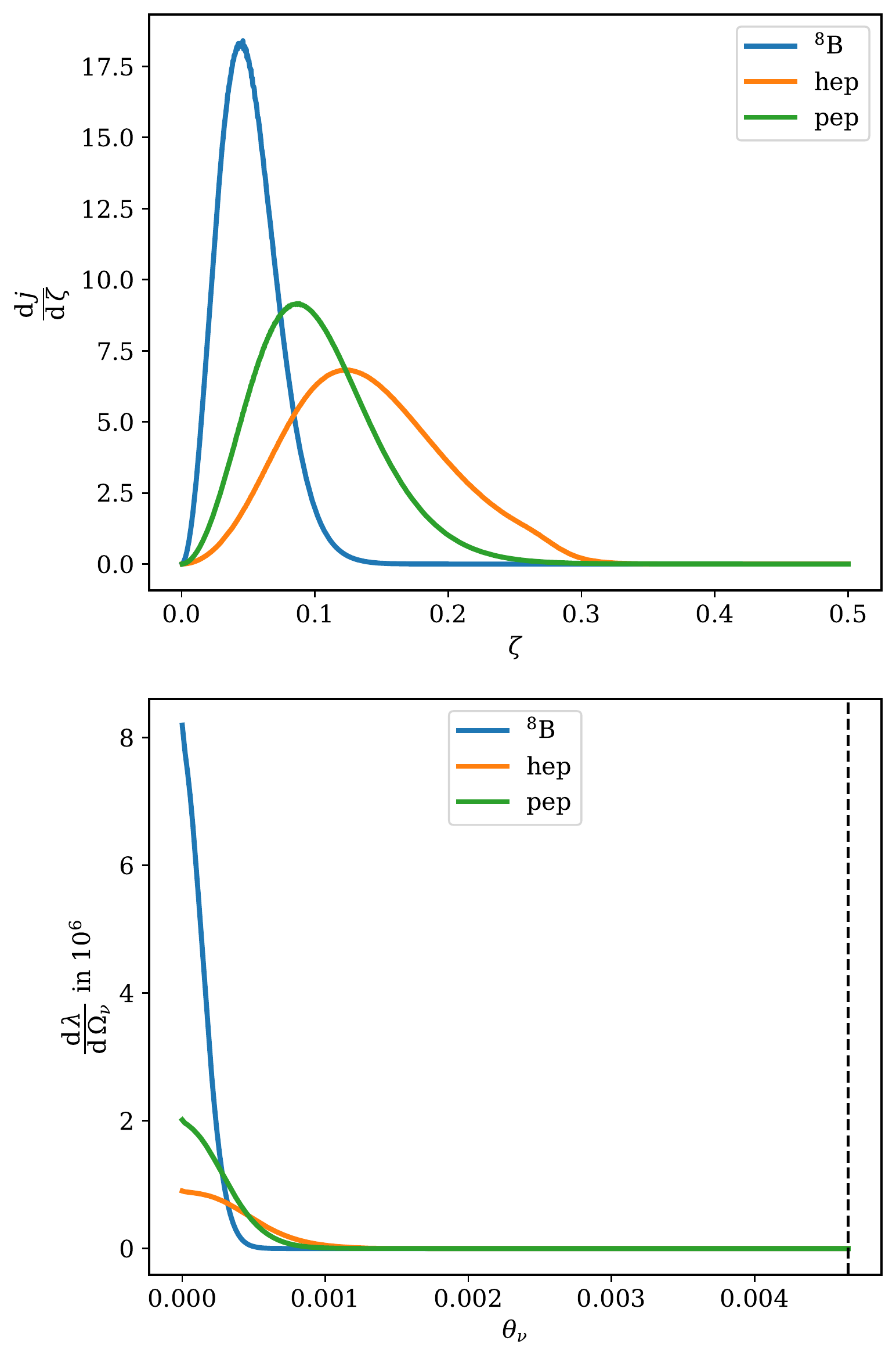}
	\caption{
		The normalised radial neutrino $^8$B, hep and pep
		production rate
		as a function of $\zeta = r/\rsun$ taken
		from the GS98 composition of \cite{Vinyoles:2016djt} on top
		and the resulting
		angular distribution as a function of $\theta_\nu$
		on the bottom. Note that we assume a radial symmetric
		distribution. The dashed vertical line in the plot
		on the bottom denotes the optical size of the sun.
		\label{fig:SolarNeutrinoLuminosityProfiles}}
\end{figure}

The resulting distributions together with the original radial distributions
for the $^8$B, the hep and the pep flux are shown in
Fig.~\ref{fig:SolarNeutrinoLuminosityProfiles}.

The most well experimentally studied neutrinos, the $^8$B neutrinos
are concentrated very much in the center of the sun.
While the optical angular diameter of the sun in the sky
is about 0.5$^\circ$ it is just about 0.07$^\circ$ for the
$^8$B neutrinos and resolving them is certainly a challenge.
At this point we would like to comment on the result in
\cite{Davis:2016hil}. From their Fig.~1 we have the impression
that their underlying neutrino angular distribution has a peak
at $\cos \theta_\nu < 1$ since their electron angular distribution
peaks at $\cos \theta_e \approx 0.984 < 1$. In our case, both neutrino
and electron angular distribution peak at the center of the sun. We will come
back to this point later.

The hep and the pep neutrinos are somewhat better to resolve
as their distributions are about two to three times wider than
the $^8$B neutrino one leading
to production zones with a diameter of about 0.14$^\circ$
which is nevertheless still small.

\begin{figure}
\centering
\includegraphics[width=0.75\textwidth]{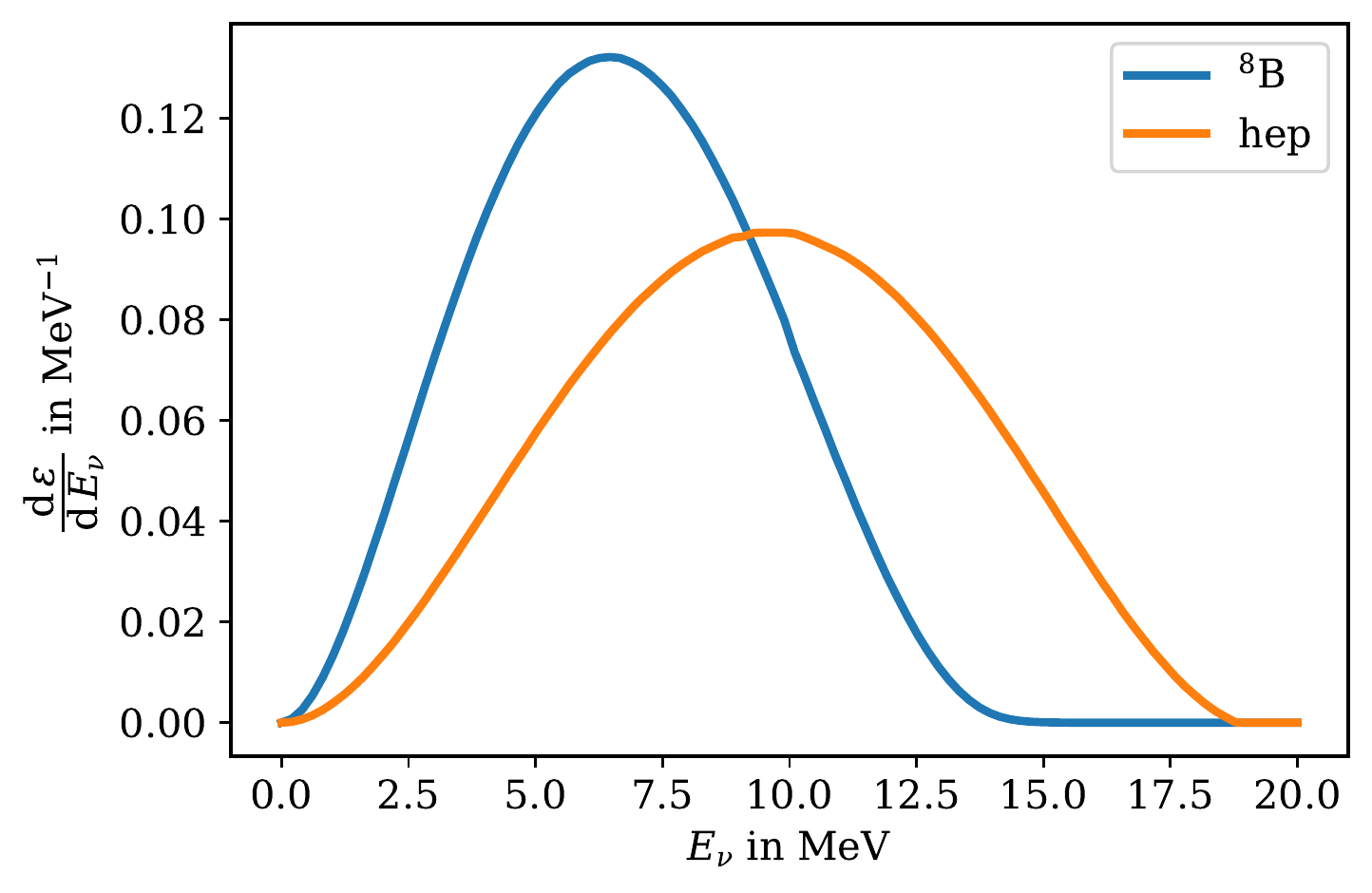}
\caption{
The normalised solar neutrino
energy spectrum for $^8$B and hep neutrinos
taken from \cite{Vitagliano:2019yzm}.
For the hep neutrinos we take the average of the two data sets
provided there.
\label{fig:SolarNeutrinoEnergyDistribution}}
\end{figure}

As a reminder before we continue, the energy distribution
of the produced neutrinos is to a good approximation independent
from the production zone and we will use the
energy distributions from \cite{Vitagliano:2019yzm}.
For completeness and convenience of the reader
we show the normalised $^8$B and hep energy distributions
in Fig.~\ref{fig:SolarNeutrinoEnergyDistribution}. The 
pep neutrinos are
monochromatic with a neutrino energy $E_\nu^{\text{pep}} = 1.445$~MeV.

\subsection{\texorpdfstring{$^{\mathbf{8}}$B Neutrinos}{8B Neutrinos}}

At this point we have all the basic ingredients collected to discuss how
we expect the $^8$B neutrinos from the sun look via electrons.

We again begin with the double differential event rate
\begin{align}
\frac{\diff R_{\text{8B}}}{\diff E_e \diff \Omega_e}  
	&= \frac{N_e \, f_0^{\text{8B}}}{2 \pi \,M_D}  \int \diff \Omega_\nu   \frac{\diff \epsilon_{\text{8B}}}{\diff E_\nu} (\bar{E}_\nu)  \frac{\diff \sigma}{\diff E_e}(E_e, \bar{E}_\nu)   \frac{ \diff \lambda_{\text{8B}}(\Omega_\nu)}{\diff \Omega_\nu} \nonumber\\
	&\phantom{= \frac{N_e \, f_0^{\text{8B}}}{2 \pi \,M_D}  \int \diff \Omega_\nu} \times \frac{\bar{E}_\nu \, \sqrt{E_e^2 - m_e^2}}{|\hat{p}_\nu \cdot \hat{q}_e \sqrt{E_e^2 - m_e^2} - (E_e - m_e)| }  \;,
\label{eq:SolarNuDoubleDifferentialRate}
\end{align}
with
\begin{equation}
	\bar{E}_\nu = \frac{m_e (E_e - m_e)}{ \hat{p}_\nu \cdot \hat{q}_e \sqrt{E_e^2 - m_e^2} - (E_e - m_e) }  
\end{equation}
and
\begin{equation}
	\hat{p}_\nu \cdot \hat{q}_e = \sin \theta_\nu \cos \phi_\nu \cos \phi_e \sin \theta_e + \sin \theta_\nu \sin \phi_\nu \sin \phi_e \sin \theta_e  + \cos \theta_\nu \cos \theta_e \;. 
\end{equation}
The remaining integration over $\Omega_\nu$ is 
highly non-trivial and has to be evaluated numerically. For some comments
on the derivation of these formulas we refer to App.~\ref{app:SolarNu}.
The energy distribution and flux factor
$f_0^{\text{8B}} = 5.46 \times 10^6$~cm$^{-2}$~s$^{-1}$
is the GS98 result from \cite{Vitagliano:2019yzm}.

Like in the previous section we will also use here a constant electron
neutrino survival probability $P_{ee} = 0.37$ \cite{BOREXINO:2018ohr} and use
for
the differential cross section in the above formula
\begin{equation}
	\frac{\diff \sigma (E_e, E_\nu)}{\diff E_e} = P_{ee}(E_\nu) \frac{\diff \sigma (\nu_e e \to \nu_e e)}{\diff E_e} + (1 - P_{ee}(E_\nu)) \frac{\diff \sigma (\nu_l e \to \nu_l e)}{\diff E_e} \;.
\end{equation}

\begin{figure}
	\centering
	\includegraphics[width=0.9\linewidth]{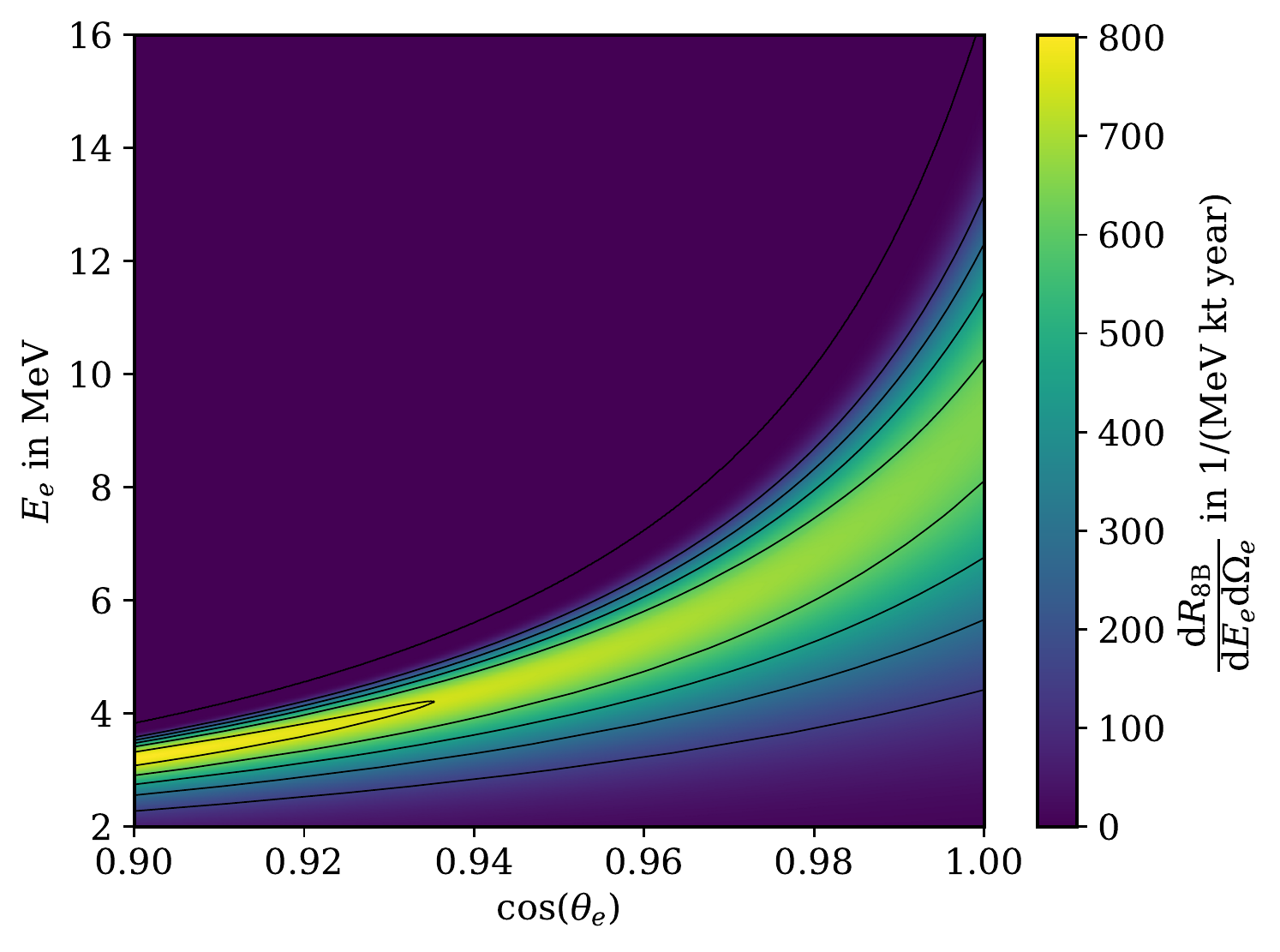}
	\caption{
		Double differential distribution of electrons recoiling from scattering with solar $^8$B neutrinos for a water Cherenkov detector. We show the distribution for $\phi_e = 0$.
		\label{fig:8B_dRdEedOmegae}}
\end{figure}
	
In Fig.~\ref{fig:8B_dRdEedOmegae} we show the double differential rate
in the $\cos \theta_e$-$E_e$ plane. We see that this picture is hard to distinguish
from the case of the point or ring source shown in Fig.~\ref{fig:8B_Comparison}
since the distribution
is extremely narrow in the sky.
Again the structure here is a non-trivial overlay of different
neutrino energies originating from different places within the sun. Although the dominant
part is clearly coming from the fact that for a given scattering angle different
neutrino energies contribute with different cross sections and hence rate which is folded
with how likely that energy is.

\begin{figure}
	\centering
	\includegraphics[width=0.75\linewidth]{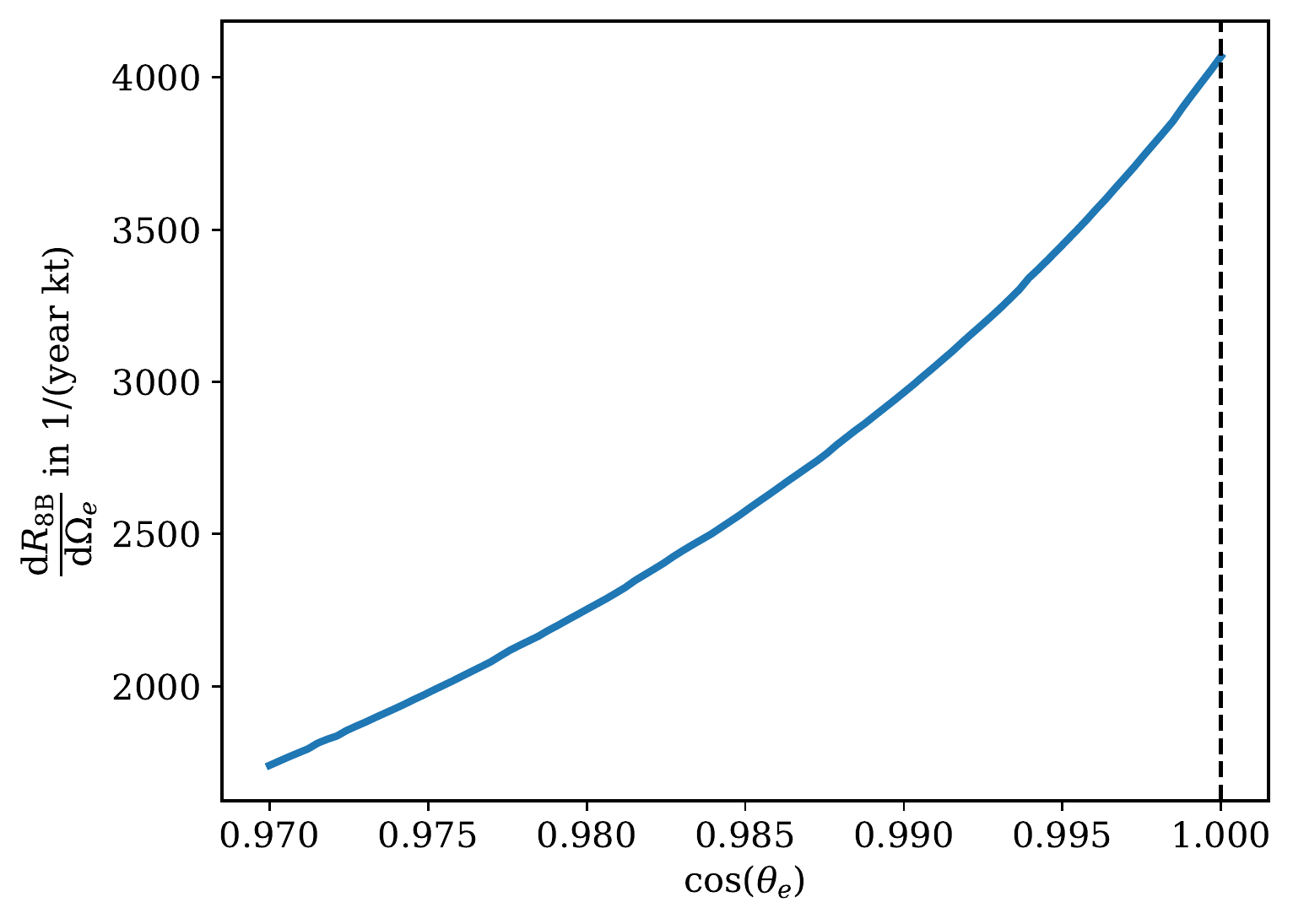}
	\caption{
		Angular distribution of electrons scattered by solar $^8$B neutrinos
		for a water Cherenkov detector. We show the distribution
		for $\phi_e = 0$ and stress that the distribution is independent of $\phi_e$.
		We integrated the electron recoil energy from $T_e = 3.5$~MeV to the maximally  possible energy.
		The dashed vertical line in the plot denotes the optical size of the sun.
		\label{fig:8B_dRdOmegae}}
\end{figure}

We like to briefly comment on the angular distribution more often seen in the literature.
We show such a distribution in Fig.~\ref{fig:8B_dRdOmegae}, which is obtained by integrating
eq.~\eqref{eq:SolarNuDoubleDifferentialRate} over the electron energy.
The event number in this figure is calculated with respect to the Super-Kamiokande detector, i.e.,
a kinetic energy threshold $T_e \geq 3.5$~MeV, cf.~\cite{Sekiya:2016nnn}, is taken for the water target. Since the distribution is
radially symmetric,
it is sufficient to plot it for a fixed value of $\phi_e = 0$.

What we see here is a rather featureless distribution which has its maximum
at $\cos \theta_e = 1$ and then smoothly falls for smaller values
of $\cos \theta_e$ which agrees, for instance, with the data shown in
\cite{Super-Kamiokande:2016yck}.
This distribution should be equivalent to the distribution shown
in Fig.~1 of \cite{Davis:2016hil} apart from using different units.
While they obtain a pronounced peak at $\cos \theta_e \approx 0.984$
we do not find such a peak in our results.
We checked that this remains true also if we
use the higher threshold of $T_e = 5$~MeV
of \cite{Davis:2016hil}. Nevertheless, we can understand our
result as the peak of the electron angular distribution
falls together with the peak of the neutrino luminosity
distribution. Unfortunately, \cite{Davis:2016hil} does not
show the assumed angular distribution of neutrino luminosity,
which would be very useful for comparison. It might also be
that they included energy dependent efficiency factors and other
uncertainties, which we do not know at this point.

\subsection{hep Neutrinos}

The second example we like to discuss is the double differential rate of
the so-called hep neutrinos.
Compared to $^8$B neutrinos, hep neutrinos can reach a higher maximal energy, which is
almost $19$ MeV while the maximal energy of $^8$B neutrinos is about 16 MeV.
The hep neutrinos are less studied experimentally because their flux is rather suppressed
compared to that of $^8$B neutrinos. The GS98 prediction 
for the flux of hep neutrinos is $0.8 \times 10^4$~cm$^{-2}$~s$^{-1}$ compared
to $5.46 \times 10^6$~cm$^{-2}$~s$^{-1}$ for the $^8$B neutrinos \cite{Vitagliano:2019yzm}.

So apart from the magnitudes and spectral shapes of the fluxes, the physics for
$^8$B and hep neutrinos is essentially the same and we shall use the same
formulas as in the previous section. To be specific, we take
$f_0^{\text{hep}} = 0.8 \times 10^4$~cm$^{-2}$~s$^{-1}$.
The production zones of hep neutrinos are taken from
\cite{Vinyoles:2016djt} resulting in the angular distribution derived
at the beginning of this section, cf.~Fig.~\ref{fig:SolarNeutrinoLuminosityProfiles},
and for the energy distribution we use the average
of the quoted minimum and maximum from \cite{Vitagliano:2019yzm} and normalise
the distribution.
The interaction cross section between hep neutrinos and electrons is taken to
be the same as the one in the case of $^8$B
neutrinos.

\begin{figure}
	\centering
	\includegraphics[width=0.9\linewidth]{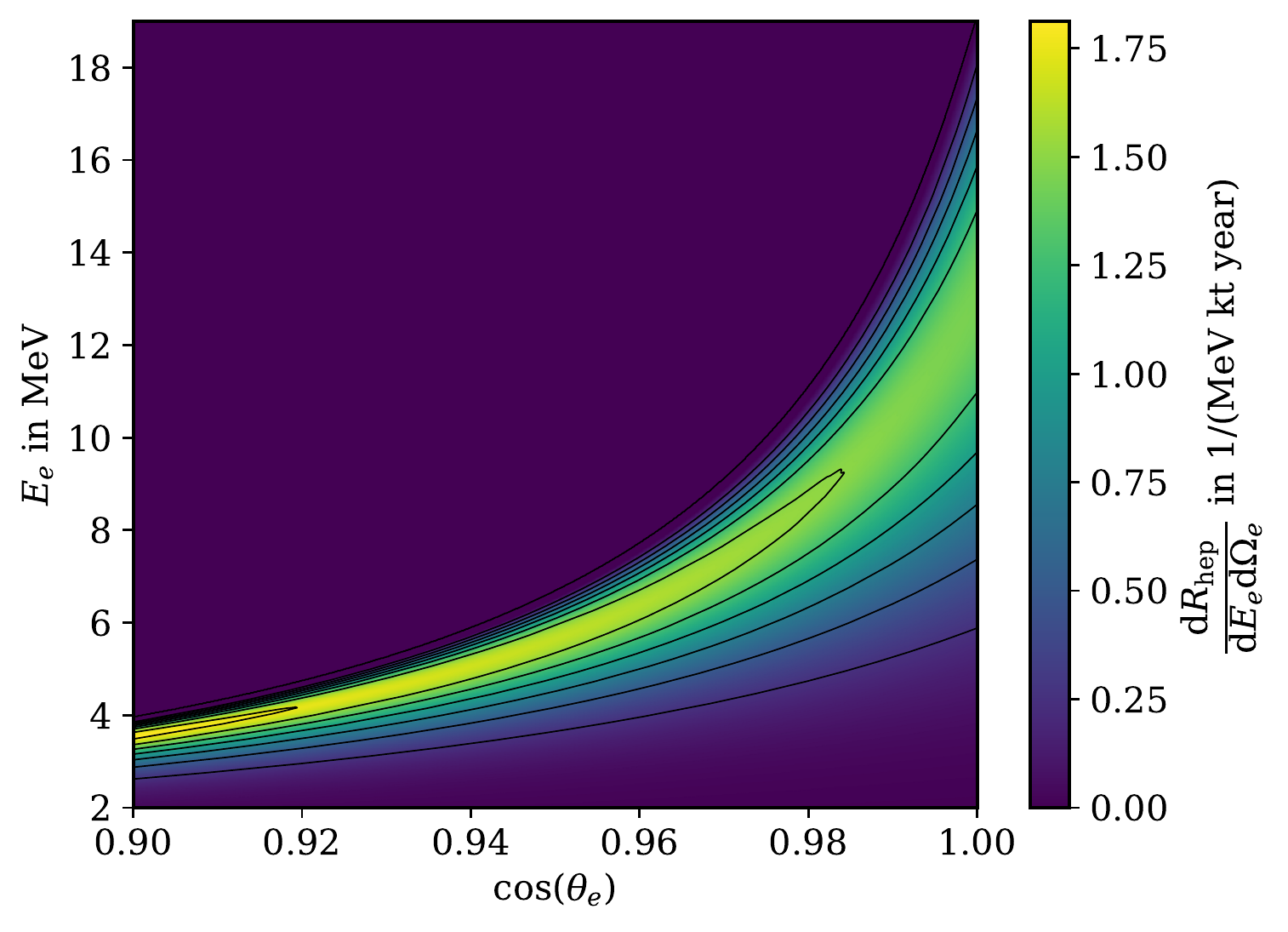}
	\caption{
		Double differential distribution of electrons recoiling from scattering with solar hep neutrinos
		for a water Cherenkov detector. We show the distribution for $\phi_e = 0$.
		\label{fig:hep_dRdEedOmegae}}
\end{figure}

We then find for the double differential distribution the result shown
in Fig.~\ref{fig:hep_dRdEedOmegae}. Not surprisingly, shapes of the contours
resemble those of $^8$B. The most notable differences is the larger
possible electron energies and the much lower event rate compared to $^8$B.

\begin{figure}
	\centering
	\includegraphics[width=0.75\linewidth]{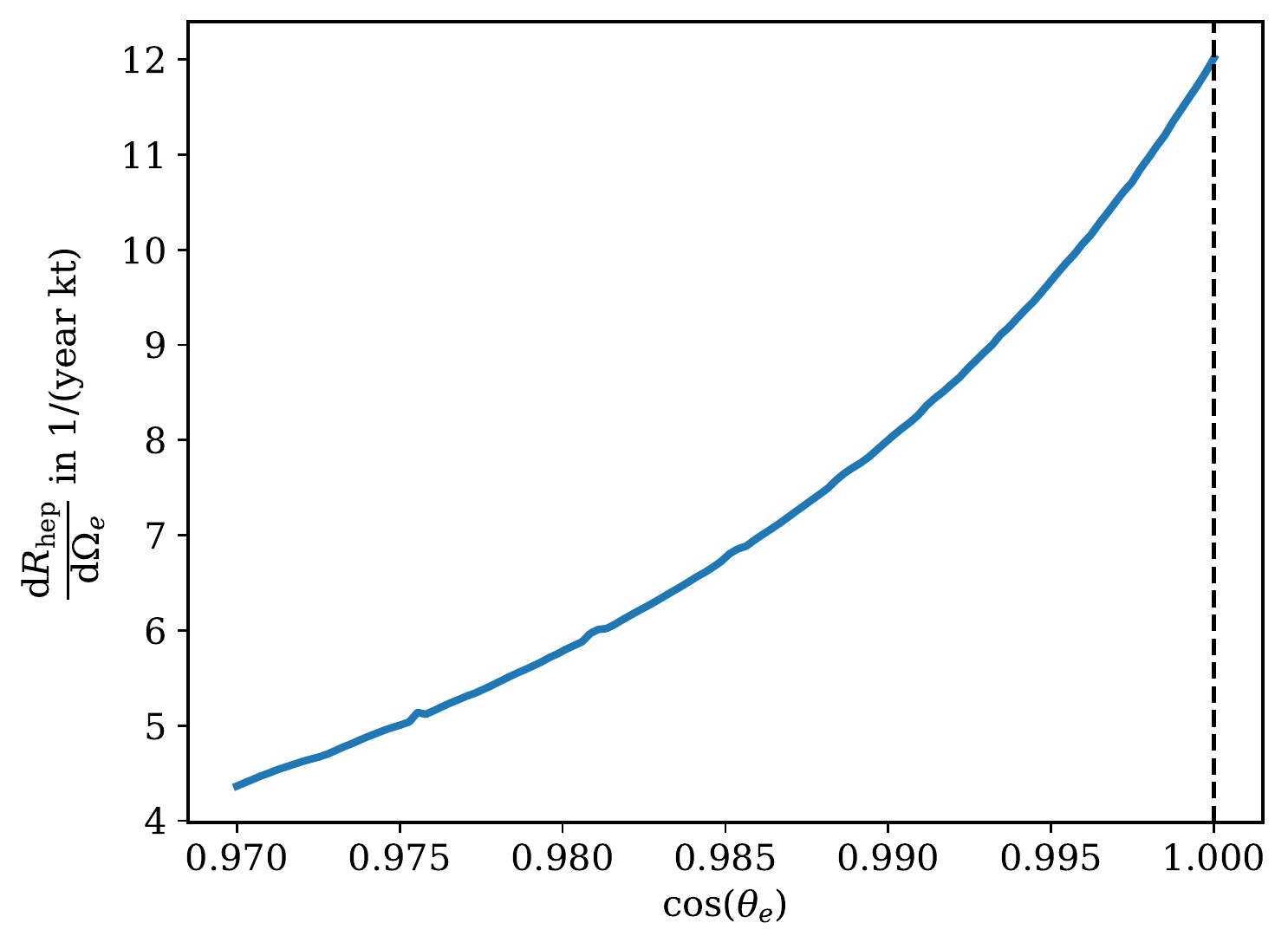}
	\caption{
		Angular distribution of electrons scattered by solar hep neutrinos
		for a water Cherenkov detector. We show the distribution
		for $\phi_e = 0$ and stress that the distribution is independent of $\phi_e$.
		We integrated the electron recoil energy from $T_e = 3.5$~MeV to the maximally possible energy.
		The dashed vertical line in the plot denotes again the optical
		size of the sun.
		\label{fig:hep_dRdOmegae}}
\end{figure}

We also show the angular distribution for hep neutrinos in
Fig.~\ref{fig:hep_dRdOmegae}. As before for the $^8$B neutrinos,
cf.\ Fig.~\ref{fig:8B_dRdOmegae} the distribution is very featureless.
The small bumps are due to numerics  and do not correspond
to any identifiable physical features. That is also consistent with our
expectations as discussed before.

\subsection{pep Neutrinos}

The third example we like to discuss is the so-called pep neutrinos.
Different from $^8$B and hep neutrinos, pep neutrinos are monochromatic. 
In principle, this could make the reconstruction of source geometry easier.

On the other hand it is quite challenging to detect pep neutrinos due to their relatively
low energy of only $E_\nu^{\text{pep}} = 1.445$~MeV, although their
total flux on Earth is not
very small. The GS98 prediction gives
$f_0^{\text{pep}} = 1.44 \times 10^8$~cm$^{-2}$~s$^{-1}$~\cite{Vitagliano:2019yzm}, which is significantly larger than the $^8$B neutrino flux.

Again we just quote here the results for the distributions of the event rate
and refer the interested reader to App.~\ref{app:SolarNu} for a detailed derivation.
For the double differential rate we find
\begin{align}
\frac{\diff R_\text{pep}}{\diff E_e \diff \Omega_e}&= 
\frac{N_e \, f_0^{\text{pep}}}{ \pi \,M_D}   \frac{\diff \sigma (E_e, E_\nu^{\text{pep}})}{\diff E_e}  \int \diff \cos \theta_\nu \frac{ \diff \lambda_{\text{pep}}(\cos \theta_\nu)}{\diff \Omega_\nu} \,  \nonumber\\
&\quad \times  \left( \sin^2 \theta_\nu \, \sin^2 \theta_e - \left(\frac{(E_e - m_e)(E_\nu^{\text{pep}} + m_e) }{E_\nu^{\text{pep}} \, \sqrt{E_e^2 - m_e^2}} - \cos \theta_e \cos \theta_\nu \right)^2   \right)^{-\tfrac{1}{2}} \;.
\end{align}
So we are just left with the integration over $\cos \theta_\nu$,
which has to be evaluated numerically. To obtain only the physical solution,
we have to ensure that the term under the square root remains positive.

\begin{figure}
	\centering
	\includegraphics[width=0.9\linewidth]{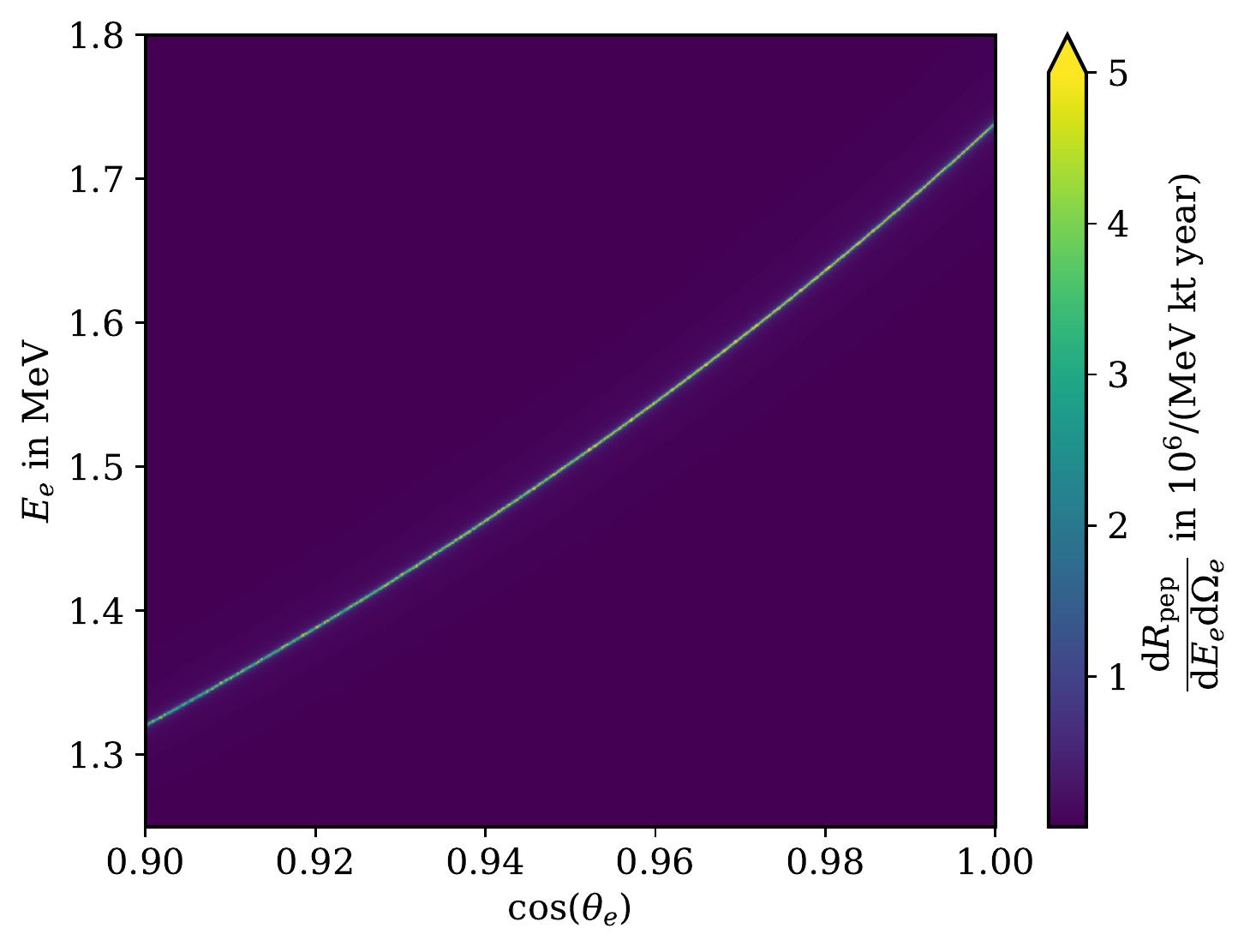}
	\caption{
		Double differential distribution of electrons recoiling from scattering with solar pep neutrinos. We show the distribution for $\phi_e = 0$.
		To increase the visibility of the extremely thin distribution
		we cap the color scale at $5 \times 10^6$ 1/(MeV~kt~year).
		\label{fig:pep_dRdEedOmegae}}
\end{figure}

We show the resulting double differential rate in Fig.~\ref{fig:pep_dRdEedOmegae}.
As expected from what we have seen in Sec.~\ref{sec:Simple}
and taking into account the rather low neutrino energy and narrow angular distribution,
the distribution looks quite similar to a line although it has a finite width.
Resolving this in an experiment might be quite challenging. On the other hand,
this strong correlation could be exploited to distinguish a pep neutrino signal
from background sources.

Since the pep signal is below typical thresholds for water Cherenkov detectors,
we do not show the angular distribution of the event rate
here.

\section{Impacts of Angular and Energy Uncertainties in Experiments}
\label{sec:ExperimentalErrors}

So far we have treated everything from a purely theoretical viewpoint.
Particularly, we have not included reconstruction uncertainties in energies and directions of electrons. 
Such uncertainties lead to an additional
blurring of the picture. One would also need to consider
how experimental errors affect the Radon transform which allows
the reconstruction of the neutrino image from electron data, cf.~Sec.~\ref{sec:Framework},
which we leave for future works.

To illustrate the impact of experimental uncertainties, we assume in this section
a very simple Gaussian error model for the
reconstructed energy in the detector, $E_d$, and direction
$\hat{q}_d$. Following \cite{Beacom:1998fj}, we assume that the above two
uncertainties are independent from each other. Hence the reconstructed
double differential rate is given by
\begin{align}
\label{eq:Smearing}
\frac{\diff R}{\diff E_d \diff \Omega_d} = \int \diff E_e \diff \Omega_e \frac{\diff R}{\diff E_e \diff \Omega_e} f_E(E_e,E_d,\sigma_E) f_\theta(\Omega_e,\Omega_d,\sigma_\theta) \;.
\end{align}
The functions $f_E$ and $f_\theta$ are Gaussian error functions
given by
\begin{align}
f_E = \mathcal{N}_E \exp\left( - \frac{(E_e - E_d)^2}{2 \, \sigma_E^2}\right) \;, \\
f_\theta = \mathcal{N}_\theta \exp\left( - \frac{\arccos^2(\hat{q}_e \cdot \hat{q}_d) }{2 \, \sigma_\theta^2}\right) \;,
\end{align}
where $\mathcal{N}_E$, $\mathcal{N}_\theta$ are normalisation constants,
for more details, see App.~\ref{app:Errors}. We want to stress here that this is
strongly simplified and should only serve some simple illustrative purposes
to see, if experimental uncertainties have the tendency to make the distinction
between different images more difficult.

\begin{figure}
\centering
\includegraphics[width=0.8\linewidth]{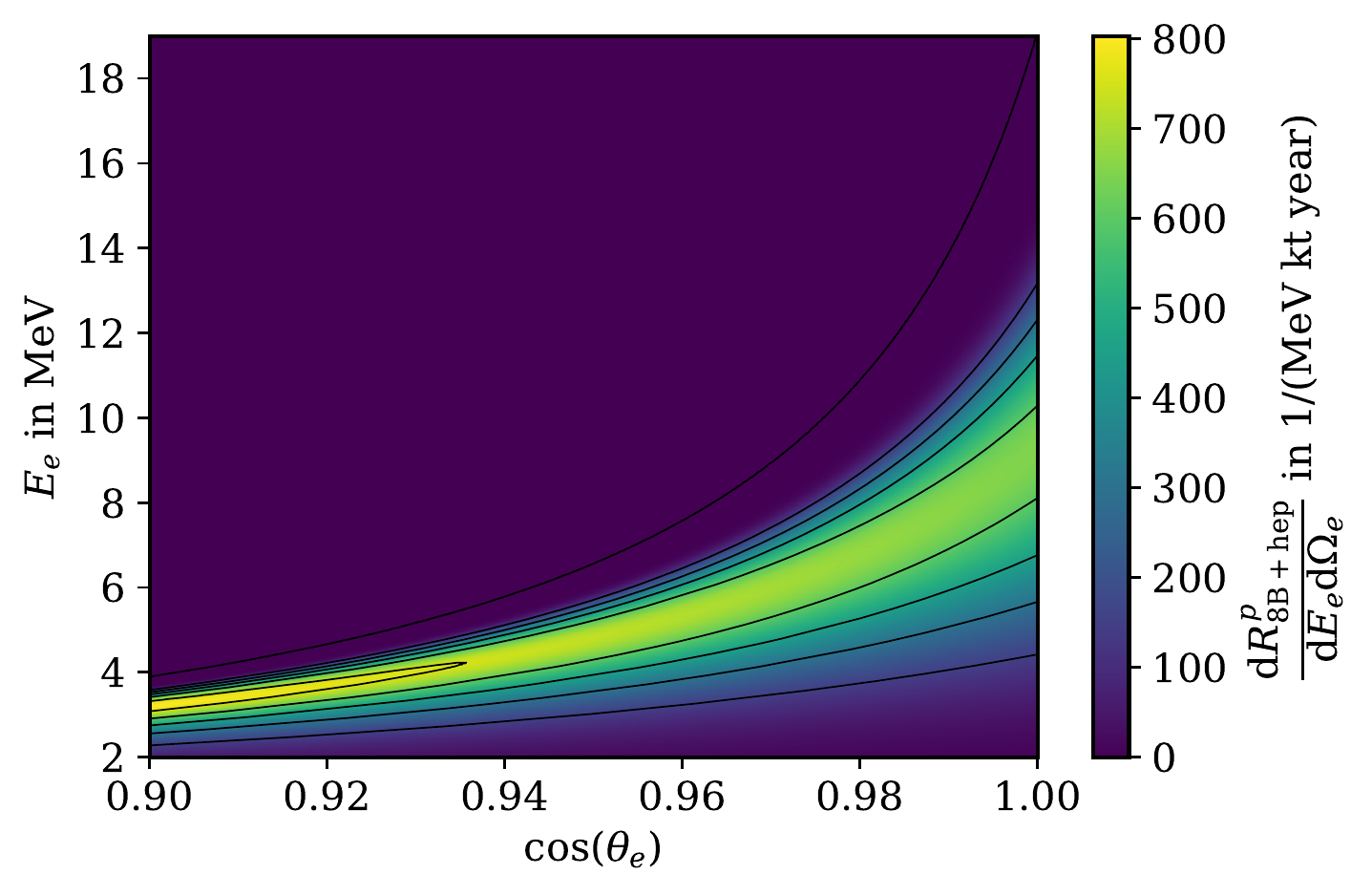} 
\includegraphics[width=0.8\linewidth]{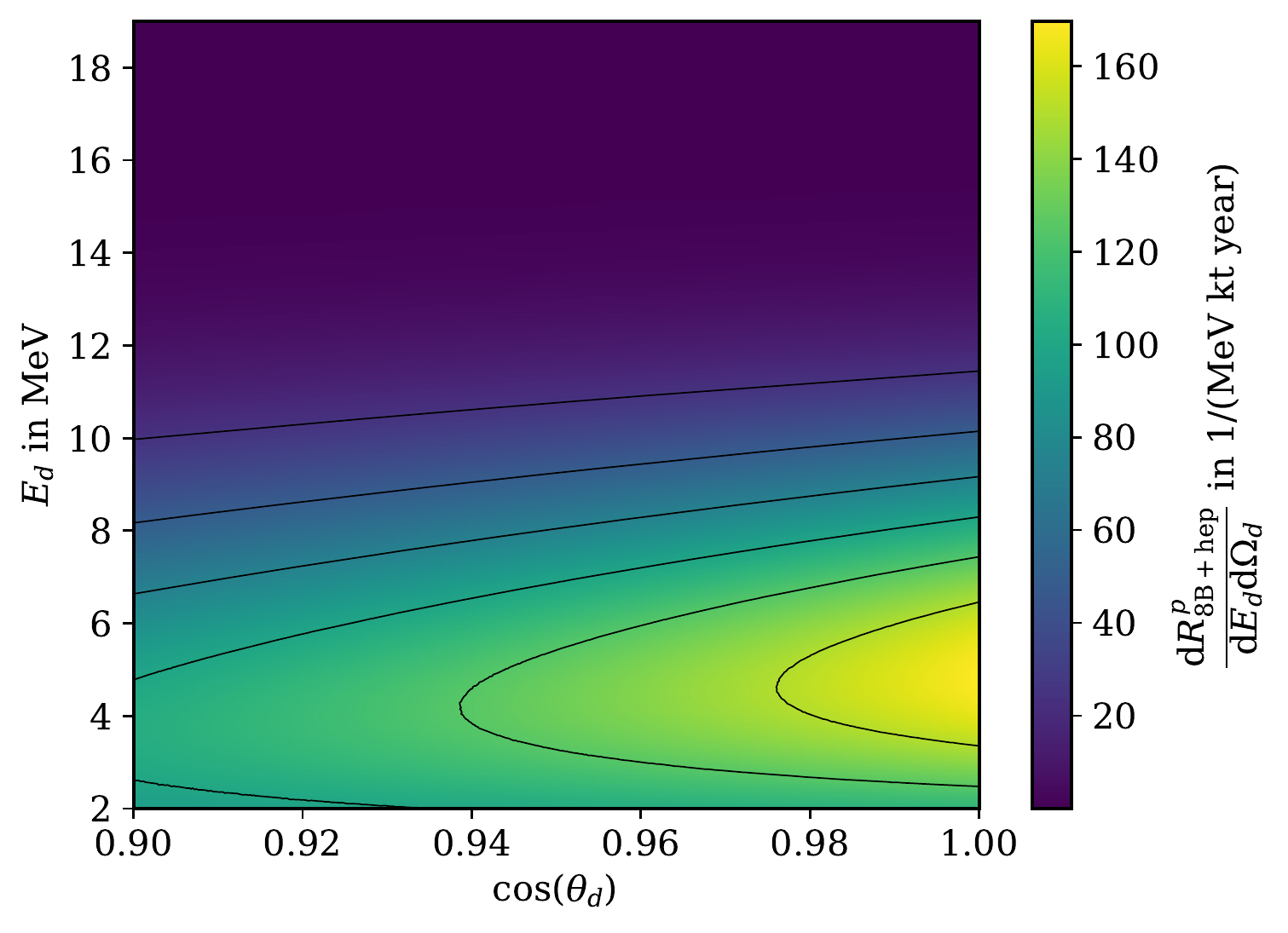}
\caption{
The double differential rate for a point source with
$^8$B and hep neutrino
fluxes.
On top we show the
original distribution while on the bottom we show the distribution
after the inclusion of simplified experimental smearing.
We show the distributions for $\phi_{e/d} = 0$.
\label{fig:8B_hep_Point}}
\end{figure}

We then calculated the double differential rate after doing this additional
smearing, see~Fig.~\ref{fig:8B_hep_Point}, where we assumed a point source and included
$^8$B and hep neutrino fluxes. For an immediate comparison, we show both  double differential distributions
before and after smearing in the plot. For the energy resolution we assumed $\sigma_E = 1.5$~MeV
and for the angular resolution we have set $\sigma_\theta = 20^\circ$ which are typical
numbers for a water Cherenkov detector, cf.~\cite{Suzuki:2019jby}.
We note that the hep neutrino contribution is
hard to see by comparing this result to Fig.~\ref{fig:8B_dRdEedOmegae}. It is
not surprising since hep neutrino flux and hence the expected rate is so much lower
than the $^8$B component, cf.~Fig.~\ref{fig:hep_dRdEedOmegae}, despite
that the hep neutrino contribution dominates for electron energies beyond 16 MeV. 

The smeared distribution looks substantially different from the original
picture which
is not surprising. The horizontal axis corresponds to a range for
$\theta_d$ of about 0$^\circ$-26$^\circ$. So it is not even
two $\sigma_\theta$ wide. The smearing due to the energy resolution
on the other hand is comparatively small. Hence, we consider
it more urgent to improve the angular resolution should one want to
resolve structures in the neutrino picture of the sun.

To understand this prospects better we also computed 
for comparison the smeared distribution for a ring source
the size of the sun and compared it to the displayed point source,
as we did in Sec.~\ref{sec:PointvsRing}. Nowhere
in the displayed region we found a difference between the two distributions larger than
1\%, which is different from comparing two sources 
without taking into account the smearing.
Therefore we expect that experimental uncertainties can blur the neutrino images as we have seen here. In fact the detection efficiency due to event selection cuts will reduce the statistics and consequently further blur the neutrino images.  
We leave detailed considerations such as the above for future
publications.

Similar to Sec.~\ref{sec:SunInNeutrinos}, one can calculate the double differential
rate for a realistic $^8$B neutrino source with smearing effects considered. However,
it is not particularly illuminating to show the result here since it is not visually
distinguishable from the result of a point source.

At this point one might dismiss our exercise here as unfeasible. In fact,
it seems far fetched to measure structures which are smaller than 0.1$^\circ$
with an instrument which has a 20$^\circ$ resolution. But this is a bit too simplistic since
the 20$^\circ$ resolution applies to an individual electron measurement. Would we suppose
all electrons come from a single point source and the error is Gaussian we could determine
the position of the source with a resolution $\sigma_\theta/\sqrt{N_{\text{evt}}}$, where
$N_{\text{evt}}$ is the event number. Super-Kamiokande reported about 1000 solar
neutrino events in phase IV~\cite{Super-Kamiokande:2016yck} which would lead to a
naive angular resolution of the right order. Similar arguments were used 
in \cite{Davis:2016hil}. Hence, in the future with significantly increased statistics
it might be possible to provide an upper bound on the size of the sun in neutrinos
even without improving the angular resolution for an individual event drastically.

We also want to remind here, that we just use solar neutrinos as a well understood
template for our formalism and that other sources which are larger and/or
have higher energies might fare better. 
One of us, for instance, discussed a hypothetical neutrino signature from DM annihilation in
the Earth's core~\cite{Lin:2014hla}. There they found that for DM mass as heavy as $10^4$~GeV, the muon-track
resolution in IceCube and the size of the neutrino production zone are both around 1$^\circ$.
To image such a neutrino production zone, it is necessary to calculate angular distributions of muons
 resulting from deep inelastic neutrino-nucleon scatterings. Such a calculation is however beyond the scope of the current work~\cite{update_cal}.

\section{Summary and Conclusions}
\label{sec:Summary}

In this article we have investigated how one could take a neutrino ``picture''
by studying the energy and angular distributions of elastically scattered electrons.
We show how the electron distributions are related to the original
neutrino distributions and also briefly mention how this relation can be inverted.
These formulas are in fact strongly inspired by the formalism developed for directional
dark matter searches. However the formalism here is more involved due to the relativistic
nature of the incoming neutrinos.

We have applied our formalism first to simple toy examples where we fixed the neutrino
energy and assumed the source to be either point- or ring-like.
These are two extreme cases and we have discussed how much they differ in the angular
and the double differential rate distributions.
Assuming a neutrino energy spectrum similar to that of $^8$B solar neutrinos, we have seen
that the difference of the double differential rate for a point and a ring source could
be up to a few percent.
On the other hand, this occurs in regions where the event rate is rather low
making such a distinction rather challenging.

We have chosen solar neutrinos here, since they are among the most well-studied
extended neutrino sources where there is not only significant amount of experimental data
but also elaborated theoretical 
calculations predicting the neutrino production rates within the solar layers. These theoretical
predictions can be turned 
into an angular neutrino luminosity distribution which together with the predicted energy spectrum
and scattering cross sections were then translated by us into neutrino pictures of the sun seen
via electrons. We did this for the dominant $^8$B and hep neutrinos and also
commented on the pep neutrinos 
and remind that they are 
produced in different regions of the sun. By comparing double differential event
rates for these neutrinos, we found the specific ranges of electron energy for
different sources. Although the events of hep neutrinos are not
copious, they could be distinguished from $^8$B neutrinos by taking an energy cut
at $E_e\sim 16$~MeV. It is expected that hep neutrinos will be identified
by the HyperK experiment.
The pep neutrinos are interesting because they are monochromatic. However, their
energy is too low for them to be detected by current and upcoming water Cherenkov
detectors which is our
focus in this paper.

Uncertainty in reconstructing the electron direction is unfortunately not small.
Hence, we also studied
how a simple error model would affect the picture and the difference between a
point and
a ring source for a combined $^8$B and hep energy flux. As one might have expected,
the difference
is washed out by our simplified consideration of experimental uncertainties.

We envisage that the understanding of the solar model will be improved with the
measurement of hep and pep neutrinos by HyperK and other future neutrino experiments.
Our results stress the importance 
of future improvements in event statistics, energy resolutions, and angular resolutions.

In conclusion,
we have established a theoretical framework for relating the neutrino
image to an image of scattered particles which can be measured directly. This framework
is not exclusively limited to solar neutrinos but can be applied to all kinds
of extended neutrino sources such as, for instance, geoneutrinos, lunar neutrinos,
neutrinos from DM interactions in various settings,
atmospheric
and solar atmospheric neutrinos or neutrinos from the galactic center just to name a few
possibilities. 
Our work is
applicable to all these cases should one like to reconstruct the neutrino picture taken
via any charged leptons or hadrons.

\section*{Acknowledgements}

We thank Meng-Ru Wu for helpful discussions.
MS is supported by the Ministry of Science and Technology (MOST) of Taiwan under grant numbers MOST 107-2112-M-007-031-MY3, MOST 110-2112-M-007-018 and MOST 111-2112-M-007-036; GLL is supported by MOST of Taiwan under grant numbers MOST 107-2119-M-009-017-MY3 and MOST 110-2112-M-A49-006; TC acknowledges the support from National Center for Theoretical Sciences.
We would like to thank the referee for useful comments.

\begin{appendix} 

\section{Elastic Neutrino Electron Scattering Cross Sections}
\label{app:CrossSections}

For the convenience of the reader we quote here
the differential elastic neutrino electron scattering cross section
which we use throughout the paper taken from \cite{Marciano:2003eq}
rewritten in terms of $E_e$
\begin{align}
\frac{\diff \sigma(\nu_\alpha e \to \nu_\alpha e)}{\diff E_e} &= \frac{2 \, G_\mu^2 m_e }{\pi}
\left[ g_{\alpha L}^2 + g_{\alpha R}^2 \frac{(E_\nu - E_e + m_e)^2}{E_\nu^2}
+  g_{\alpha L} g_{\alpha R}  \frac{m_e (E_e - m_e)}{E_\nu^2} \right] \label{eq:sigma}
\end{align}
where for 
$\alpha = \mu, \tau$, $g_{\alpha L} = \tfrac{1}{2} - s_W^2$, $g_{\alpha R} = s_W^2 \equiv \sin^2 \theta_W$ the weak mixing angle
and for $\alpha = e$, $g_{e L} = \tfrac{1}{2} + s_W^2$, $g_{e R} = - s_W^2 \equiv \sin^2 \theta_W$. $G_\mu$ is the Fermi constant.

\section{Details about the Derivations for the Double Differential Event Rates}
\label{app:RingPointApp}

In this section we collect some more details on how to derive
double differential event rates for certain setups.

\subsection{Monochromatic Point Source}
\label{app:Point}

For the monochromatic point source we assume the following angular and energy distributions
\begin{equation}
\frac{ \diff \lambda_p(\Omega_\nu)}{\diff \Omega_\nu} = \frac{1}{2 \pi } \delta(\cos \theta_\nu - 1) 
\text{ and } \frac{\diff \epsilon_p(E_\nu)}{\diff E_\nu} = \delta(E_\nu - E_0) \;.
\end{equation}
Then
\begin{equation}
\int \diff \Omega_\nu \frac{ \diff \lambda_p(\Omega_\nu)}{\diff \Omega_\nu} \delta\left( \hat{p}_\nu \cdot \hat{q}_e - \frac{(E_e - m_e)(E_\nu + m_e) }{E_\nu \, \sqrt{E_e^2 - m_e^2}}  \right) =\delta\left( \cos \theta_e - \frac{(E_e - m_e)(E_\nu + m_e) }{E_\nu \, \sqrt{E_e^2 - m_e^2}}  \right) \;,
\end{equation}
and putting this back into the key formula, eq.~\eqref{eq:KeyFormula},
and performing the trivial integration over the neutrino energy we get the
double differential event rate
\begin{align}
\frac{\diff R_p}{\diff E_e \diff \Omega_e} &= 
\frac{N_e \, f_0}{2 \pi \,M_D}  \frac{\diff \sigma}{\diff E_e} (E_e, E_\nu = E_0)  \,  \delta\left(  \cos \theta_e - \frac{(E_e - m_e)(E_0 + m_e)}{E_0 \sqrt{E_e^2 - m_e^2}}  \right) \;.
\end{align}
Here one could choose to integrate over the electron energy or
the electron angular distribution depending on what one is interested in.

\subsection{Monochromatic Ring source}
\label{app:Ring}

We now consider a monochromatic ring source, i.e.,
\begin{equation}
\frac{ \diff \lambda_r(\Omega_\nu)}{\diff \Omega_\nu} = \frac{1}{2 \pi } \delta\left(\cos \theta_\nu - c_r \right) 
\text{ and } \frac{\diff \epsilon_r(E_\nu)}{\diff E_\nu} = \delta(E_\nu - E_0) \;,
\end{equation}
where we do not yet specify the opening angle of the ring but assume it is
small, $1 \gg 1- c_r^2 > 0$.
Repeating the previous calculations as in App.~\ref{app:Point}
\begin{equation}
\begin{split}
\int \diff \Omega_\nu \frac{ \diff \lambda_r(\Omega_\nu)}{\diff \Omega_\nu} \delta\left( \hat{p}_\nu \cdot \hat{q}_e - \frac{(E_e - m_e)(E_\nu + m_e) }{E_\nu \, \sqrt{E_e^2 - m_e^2}}  \right) = \\
\int_0^{2 \pi} \diff \phi_\nu \frac{1}{2\,\pi} \, \delta\left( (\hat{p}_\nu \cdot \hat{q}_e)_r - \frac{(E_e - m_e)(E_\nu + m_e) }{E_\nu \, \sqrt{E_e^2 - m_e^2}}  \right) \;,
\end{split}
\end{equation}
where
\begin{equation}
(\hat{p}_\nu \cdot \hat{q}_e)_r = \sqrt{1-c_r^2}  \cos \phi_\nu \cos \phi_e \sin \theta_e + \sqrt{1-c_r^2} \sin \phi_\nu \sin \phi_e \sin \theta_e  + c_r \cos \theta_e \;. 
\end{equation}
Note that we are left here with the integration over $\phi_\nu$ which is different
than for the point source case.

The double differential event rate for the monochromatic ring source is then given by
\begin{align}
\frac{\diff R_r}{\diff E_e \diff \Omega_e} &= 
\frac{N_e \, f_0}{4 \pi^2 \,M_D} 
\int_0^{2 \pi} \diff \phi_\nu \,  \frac{\diff \sigma}{\diff E_e} (E_e,E_\nu = E_0) \delta\left( (\hat{p}_\nu \cdot \hat{q}_e)_r  - \frac{(E_e - m_e)(E_0 + m_e)}{E_0 \sqrt{E_e^2 - m_e^2}}  \right)  \;.
\label{eq:RingdRdOmegaeApp}
\end{align}
Since we want to study the distribution $\diff R_r/\diff E_e \diff \Omega_e$
as well
it is useful to exploit the $\delta$-distribution. To simplify the discussion
we can use the radial symmetry of the problem and
\begin{equation}
\frac{\diff R_r}{\diff E_e \diff \Omega_e}(E_e, \cos \theta_e,\phi_e) =  \frac{\diff R_r}{\diff E_e \diff \Omega_e}(E_e, \cos \theta_e, \phi_e = 0) \;.
\end{equation}
For $\phi_e = 0$ the expression for $(\hat{p}_\nu \cdot \hat{q}_e)_r$ simplifies
and
\begin{align}
\int_0^{2 \pi} & \diff \phi_\nu \, \delta\left( (\hat{p}_\nu \cdot \hat{q}_e)_r - \frac{(E_e - m_e)(E_0 + m_e) }{E_0 \, \sqrt{E_e^2 - m_e^2}}  \right) \nonumber\\
& = \int_0^{2 \pi} \diff \phi_\nu \, \frac{\delta\left(\phi_\nu - \xi_\nu \right) + \delta\left(\phi_\nu + \xi_\nu \right)}{|\sqrt{1-c_r^2}  \sin \phi_\nu \sin \theta_e| } = \frac{2}{|\sqrt{1-c_r^2}  \sin \xi_\nu \sin \theta_e|}\;,
\end{align}
where
\begin{equation}
\xi_\nu = \arccos\left( \frac{(E_e - m_e)(E_0 + m_e) }{\sqrt{1-c_r^2}  \sin \theta_e \, E_0 \, \sqrt{E_e^2 - m_e^2}} -  \frac{c_r \cos \theta_e}{\sqrt{1-c_r^2}  \sin \theta_e} \right) 
\end{equation}
and $E_e$ and $\theta_e$ must be such that $\xi_\nu$ is real.
The double differential event rate then is
\begin{equation}
\frac{\diff R_r}{\diff E_e \diff \Omega_e} =
\frac{N_e \, f_0}{ 2 \pi^2 \,M_D}  \frac{1}{|\sqrt{1-c_r^2}  \sin \xi_\nu \sin \theta_e|}
\frac{\diff \sigma}{\diff E_e} (E_e,E_\nu = E_0) \theta(1 - \cos \xi_\nu) \theta(\cos \xi_\nu + 1)
\;.
\end{equation}

If one would want to know the 
angular distribution $\diff R_r/\diff \Omega_e$
it would be easier  to start from eq.~\eqref{eq:RingdRdOmegae}
(which is identical to eq.~\eqref{eq:RingdRdOmegaeApp})
instead. We can then use the $\delta$-function to perform the energy
integration and find
\begin{align}
\frac{\diff R_r}{\diff \Omega_e} &= \int_{E_e^{\text{thr}}}^\infty \diff E_e \frac{\diff R}{\diff E_e \diff \Omega_e} \nonumber\\
&= \frac{N_e \, f_0}{4 \pi^2 \,M_D}  \int_0^{2 \pi} \diff \phi_\nu \,   \frac{\diff \sigma}{\diff E_e} (E_e = E_e^{(r)}, E_\nu 
= E_0 )  \,  \theta(E_e^{(r)} - E_e^{\text{thr}}) \nonumber\\
&\quad \times \frac{E_0 (E_e^{(r)} + m_e) \sqrt{(E_e^{(r)})^2 - m_e^2} }{ m_e (E_0 + m_e) }\;,
\end{align}
where
\begin{equation}
E_e^{(r)} = m_e \frac{ E_0^2 ( (\hat{p}_\nu \cdot \hat{q}_e)_r^2 + 1)  + 2 \, m_e \, E_0 + m_e^2 }{E_0^2 ( 1- (\hat{p}_\nu \cdot \hat{q}_e)_r^2) + 2 \, m_e \, E_0 + m_e^2} \;.
\end{equation}
So here one would still have to integrate over $\phi_\nu$ which is not trivial
since $E_e^{(r)}$ depends on $\phi_\nu$ via $(\hat{p}_\nu \cdot \hat{q}_e)_r$.

\subsection{Comparison Ring vs.\ Point Source}
\label{app:Comparison}

Here we derive the formulas used to produce Fig.~\ref{fig:8B_Comparison}.
We will do it explicitly for an $^8$B energy spectrum and flux but one could
do the same for hep neutrinos and
all the relevant quantities would just have to be replaced with their corresponding ones.
Let us start again from our key formula, cf.~\eqref{eq:KeyFormula},
\begin{align}
\frac{\diff R_{\text{8B}}}{\diff E_e \diff \Omega_e} &=  \frac{N_e \, f_0^{\text{8B}}}{2 \pi \,M_D} \int \diff E_\nu \, \frac{\diff \epsilon_{\text{8B}}(E_\nu)}{\diff E_\nu}  \frac{\diff \sigma (E_e, E_\nu)}{\diff E_e}   \,  \nonumber\\
&\quad \times \int \diff \Omega_\nu \frac{ \diff \lambda_{\text{8B}}(\Omega_\nu)}{\diff \Omega_\nu} \delta\left( \hat{p}_\nu \cdot \hat{q}_e - \frac{(E_e - m_e)(E_\nu + m_e) }{E_\nu \, \sqrt{E_e^2 - m_e^2}}  \right) \;.
\end{align}
The point source case is then easy to get using the angular distribution
\begin{equation}
   \frac{ \diff \lambda^p_{\text{8B}}(\Omega_\nu)}{\diff \Omega_\nu} = \frac{1}{2 \, \pi} \delta(\cos \theta_\nu - 1) \;.
\end{equation}
We find
\begin{align}
\frac{\diff R^p_{\text{8B}}}{\diff E_e \diff \Omega_e} &=  \frac{N_e \, f_0^{\text{8B}}}{2 \pi \,M_D} \int \diff E_\nu \, \frac{\diff \epsilon_{\text{8B}}(E_\nu)}{\diff E_\nu}  \frac{\diff \sigma (E_e, E_\nu)}{\diff E_e}  \delta\left( \cos \theta_e - \frac{(E_e - m_e)(E_\nu + m_e) }{E_\nu \, \sqrt{E_e^2 - m_e^2}}  \right) \nonumber\\
&= \frac{N_e \, f_0^{\text{8B}}}{2 \pi \,M_D} \frac{\diff \epsilon_{\text{8B}}}{\diff E_\nu}(E_\nu^p)  \frac{\diff \sigma (E_e, E_\nu^p)}{\diff E_e}
 m_e \left| \frac{(E_e - m_e) \sqrt{E_e^2 - m_e^2}}{(E_e - m_e - \cos \theta_e \sqrt{E_e^2 - m_e^2})^2} \right|\;,
\end{align}
where
\begin{equation}
    E_\nu^p = \frac{m_e (E_e - m_e)}{m_e - E_e + \cos \theta_e \sqrt{E_e^2 - m_e^2}} \;.
\end{equation}
The ring case is slightly more complicated and based on the
angular distribution
\begin{equation}
   \frac{ \diff \lambda^r_{\text{8B}}(\Omega_\nu)}{\diff \Omega_\nu} = \frac{1}{2 \, \pi} \delta(\cos \theta_\nu - c_r) \;,
\end{equation}
where we have used in Fig.~\ref{fig:8B_Comparison}
that $c_r = \cos \theta_\text{sun} \approx 1 - 10^{-6}$.
Then 
\begin{align}
\frac{\diff R^r_{\text{8B}}}{\diff E_e \diff \Omega_e} &=  \frac{N_e \, f_0^{\text{8B}}}{2 \pi \,M_D} \int \diff E_\nu \int_0^{2\pi} \diff \phi_\nu \, \frac{\diff \epsilon_{\text{8B}}(E_\nu)}{\diff E_\nu}  \frac{\diff \sigma (E_e, E_\nu)}{\diff E_e} \nonumber\\
 &\phantom{\frac{N_e \, f_0^{\text{8B}}}{2 \pi \,M_D} \int \diff E_\nu \int_0^{2\pi} \diff \phi_\nu} \times \delta\left( (\hat{p}_\nu \cdot \hat{q}_e)_r - \frac{(E_e - m_e)(E_\nu + m_e) }{E_\nu \, \sqrt{E_e^2 - m_e^2}}  \right) \nonumber\\
&= \frac{N_e \, f_0^{\text{8B}}}{4 \pi^2 \,M_D} \int_0^{2\pi} \diff \phi_\nu \frac{\diff \epsilon_{\text{8B}}}{\diff E_\nu}(E_\nu^r)  \frac{\diff \sigma (E_e, E_\nu^r)}{\diff E_e}
\nonumber\\
&\phantom{\frac{N_e \, f_0^{\text{8B}}}{4 \pi^2 \,M_D} \int_0^{2\pi} \diff \phi_\nu \frac{\diff \epsilon_{\text{8B}}}{\diff E_\nu}(E_\nu^r) }
\times
 m_e \left| \frac{(E_e - m_e) \sqrt{E_e^2 - m_e^2}}{(E_e - m_e - (\hat{p}_\nu \cdot \hat{q}_e)_r \sqrt{E_e^2 - m_e^2})^2} \right| \;,
\end{align}
where
\begin{align}
    E_\nu^r &= \frac{m_e (E_e - m_e)}{m_e - E_e + (\hat{p}_\nu \cdot \hat{q}_e)_r \sqrt{E_e^2 - m_e^2}} \;,\\
    (\hat{p}_\nu \cdot \hat{q}_e)_r &= \sqrt{1-c_r^2}  \cos \phi_\nu \cos \phi_e \sin \theta_e + \sqrt{1-c_r^2} \sin \phi_\nu \sin \phi_e \sin \theta_e  + c_r \cos \theta_e \;.
\end{align}

\subsection{Solar Neutrinos}
\label{app:SolarNu}

In this subsection we provide more details on the derivations of the relevant formulas in
Sec.~\ref{sec:SunInNeutrinos}. We begin with the $^8$B neutrinos and start again
from our key formula, cf.~\eqref{eq:KeyFormula},
\begin{align}
	\frac{\diff R_{\text{8B}}}{\diff E_e \diff \Omega_e} &=  \frac{N_e \, f_0^{\text{8B}}}{2 \pi \,M_D} \int \diff E_\nu \, \frac{\diff \epsilon_{\text{8B}}(E_\nu)}{\diff E_\nu}  \frac{\diff \sigma (E_e, E_\nu)}{\diff E_e}   \,  \nonumber\\
	&\quad \times \int \diff \Omega_\nu \frac{ \diff \lambda_{\text{8B}}(\Omega_\nu)}{\diff \Omega_\nu} \delta\left( \hat{p}_\nu \cdot \hat{q}_e - \frac{(E_e - m_e)(E_\nu + m_e) }{E_\nu \, \sqrt{E_e^2 - m_e^2}}  \right) \;.
\end{align}
For $^8$B neutrinos both angular and energy distribution are non-trivial
and in particular there is no additional $\delta$-function to exploit.
The only $\delta$-function present can be used to evaluate the
integration over the neutrino energy and we get
\begin{align}
	\frac{\diff R_{\text{8B}}}{\diff E_e \diff \Omega_e}  
	&= \frac{N_e \, f_0^{\text{8B}}}{2 \pi \,M_D}  \int \diff \Omega_\nu   \frac{\diff \epsilon_{\text{8B}}}{\diff E_\nu} (\bar{E}_\nu)  \frac{\diff \sigma}{\diff E_e}(E_e, \bar{E}_\nu)   \frac{ \diff \lambda_{\text{8B}}(\Omega_\nu)}{\diff \Omega_\nu} \nonumber\\
	&\phantom{= \frac{N_e \, f_0^{\text{8B}}}{2 \pi \,M_D}  \int \diff \Omega_\nu} \times \frac{\bar{E}_\nu \, \sqrt{E_e^2 - m_e^2}}{|\hat{p}_\nu \cdot \hat{q}_e \sqrt{E_e^2 - m_e^2} - (E_e - m_e)| }  \;,
\end{align}
with
\begin{equation}
	\bar{E}_\nu = \frac{m_e (E_e - m_e)}{ \hat{p}_\nu \cdot \hat{q}_e \sqrt{E_e^2 - m_e^2} - (E_e - m_e) }  
\end{equation}
and
\begin{equation}
	\hat{p}_\nu \cdot \hat{q}_e = \sin \theta_\nu \cos \phi_\nu \cos \phi_e \sin \theta_e + \sin \theta_\nu \sin \phi_\nu \sin \phi_e \sin \theta_e  + \cos \theta_\nu \cos \theta_e \;. 
\end{equation}
The remaining integration over $\Omega_\nu$ is unfortunately
highly non-trivial and has to be done numerically.
	
Please note that for the  differential cross section we use 
\begin{equation}
	\frac{\diff \sigma (E_e, E_\nu)}{\diff E_e} = P_{ee}(E_\nu) \frac{\diff \sigma (\nu_e e \to \nu_e e)}{\diff E_e} + (1 - P_{ee}(E_\nu)) \frac{\diff \sigma (\nu_l e \to \nu_l e)}{\diff E_e} \;,
\end{equation}
where we have introduced the electron neutrino survival probability $P_{ee}$
which we set to $0.37$ in our numerical calculations \cite{BOREXINO:2018ohr}.

The case of hep neutrinos is formally the same. We just have to replace
the neutrino energy and angular distributions and the total flux factor
with the hep quantities in the above formulas.

The pep neutrinos are different though due to their fixed energy.
Once again beginning in our key formula we plug in the 
pep neutrino energy distribution
\begin{equation}
    \frac{\diff \epsilon_{\text{pep}}(E_\nu)}{\diff E_\nu} = \delta(E_\nu - E_\nu^{\text{pep}}) \;.
\end{equation}
Therefore, we can use this $\delta$-function to do the neutrino energy
integration and find
\begin{align}
\frac{\diff R_\text{pep}}{\diff E_e \diff \Omega_e} &=  \frac{N_e \, f_0^{\text{pep}}}{2 \pi \,M_D}   \frac{\diff \sigma (E_e, E_\nu^{\text{pep}})}{\diff E_e}   \,  \nonumber\\
&\quad \times \int \diff \Omega_\nu \frac{ \diff \lambda_{\text{pep}}(\Omega_\nu)}{\diff \Omega_\nu} \delta\left( \hat{p}_\nu \cdot \hat{q}_e - \frac{(E_e - m_e)(E_\nu^{\text{pep}} + m_e) }{E_\nu^{\text{pep}} \, \sqrt{E_e^2 - m_e^2}}  \right) \;.
\end{align}
Before we exploit the remaining $\delta$-function we want to remind
that the problem has a radial symmetry. That means the result should
not depend on $\phi_e$ and 
\begin{align}
\frac{\diff R_\text{pep}}{\diff E_e \diff \Omega_e}&(\cos \theta_e, \phi_e) = \frac{\diff R_\text{pep}}{\diff E_e \diff \Omega_e}(\cos \theta_e, 0) =  \frac{N_e \, f_0^{\text{pep}}}{2 \pi \,M_D}   \frac{\diff \sigma (E_e, E_\nu^{\text{pep}})}{\diff E_e}   \,  \nonumber\\
&\quad \times \int \diff \Omega_\nu \frac{ \diff \lambda_{\text{pep}}(\Omega_\nu)}{\diff \Omega_\nu} \delta\left( \hat{p}_\nu \cdot \hat{q}_e(\phi_e = 0) - \frac{(E_e - m_e)(E_\nu^{\text{pep}} + m_e) }{E_\nu^{\text{pep}} \, \sqrt{E_e^2 - m_e^2}}  \right) \nonumber\\
&= \frac{N_e \, f_0^{\text{pep}}}{ \pi \,M_D}   \frac{\diff \sigma (E_e, E_\nu^{\text{pep}})}{\diff E_e}  \int \diff \cos \theta_\nu \frac{ \diff \lambda_{\text{pep}}(\cos \theta_\nu,\phi_\nu^{\text{pep}})}{\diff \Omega_\nu} \,  \nonumber\\
&\quad \times  \left( \sin^2 \theta_\nu \, \sin^2 \theta_e - \left(\frac{(E_e - m_e)(E_\nu^{\text{pep}} + m_e) }{E_\nu^{\text{pep}} \, \sqrt{E_e^2 - m_e^2}} - \cos \theta_e \cos \theta_\nu \right)^2   \right)^{-\tfrac{1}{2}} \;,
\end{align}
where $\phi_\nu^{\text{pep}}$ is one of the roots of the argument of
the $\delta$-function. There are two roots $\phi_\nu^{\text{pep}} = \pm \arccos(\ldots)$
but they both give the same final result introducing a factor two.
But since $\diff \lambda / \diff \Omega_\nu$ does not depend on $\phi_\nu$
in the examples we study we can just ignore that dependence and find
\begin{align}
\frac{\diff R_\text{pep}}{\diff E_e \diff \Omega_e}&= 
\frac{N_e \, f_0^{\text{pep}}}{ \pi \,M_D}   \frac{\diff \sigma (E_e, E_\nu^{\text{pep}})}{\diff E_e}  \int \diff \cos \theta_\nu \frac{ \diff \lambda_{\text{pep}}(\cos \theta_\nu)}{\diff \Omega_\nu} \,  \nonumber\\
&\quad \times  \left( \sin^2 \theta_\nu \, \sin^2 \theta_e - \left(\frac{(E_e - m_e)(E_\nu^{\text{pep}} + m_e) }{E_\nu^{\text{pep}} \, \sqrt{E_e^2 - m_e^2}} - \cos \theta_e \cos \theta_\nu \right)^2   \right)^{-\tfrac{1}{2}} \;.
\end{align}
So we are just left with the integration over $\cos \theta_\nu$
which we have to evaluate numerically. To only get physical solutions
we also have to make
sure that the term under the square root remains positive.

\section{Details on the Approximate Error Functions}
\label{app:Errors}

Here we want to show and derive explicit expressions for the
normalisation constants used for the approximate modelling of
experimental errors in Sec.~\ref{sec:ExperimentalErrors} in
the functions
\begin{align}
f_E = \mathcal{N}_E \exp\left( - \frac{(E_e - E_d)^2}{2 \, \sigma_E^2}\right) \;, \\
f_\theta = \mathcal{N}_\theta \exp\left( - \frac{\arccos^2(\hat{q}_e \cdot \hat{q}_d) }{2 \, \sigma_\theta^2}\right) \;.
\end{align}
The explicit expressions are
\begin{align}
\int_{m_e}^{\infty} \diff E_e \, f_E \stackrel{!}{=} 1 \Rightarrow \mathcal{N}_E &= \frac{\sqrt{2}}{\sigma_E \sqrt{\pi} (1 + \erf( (E_d - m_e)/(\sqrt{2} \, \sigma_E) ) ) } \;, \\
\int \diff \Omega_e \, f_\theta \stackrel{!}{=} 1 \Rightarrow
\mathcal{N}_\theta &= \ci \sqrt{\frac{2}{\pi^3 \, \sigma_\theta^2 }} \frac{\exp\left(\frac{\sigma_\theta^2}{2}\right)}{2 \erf\left(\frac{\ci \sigma_\theta}{\sqrt{2}} \right)
-\erf\left(\frac{\ci \sigma_\theta^2 + \pi }{\sqrt{2}\sigma_\theta}\right)
-\erf\left(\frac{\ci \sigma_\theta^2 - \pi }{\sqrt{2} \sigma_\theta}\right)}   \;.
\end{align}
and
\begin{equation}
\hat{q}_e \cdot \hat{q}_d =  \cos \phi_e \sin \theta_e \cos \phi_d \sin \theta_d + \sin \phi_e \sin \theta_e \sin \phi_d \sin \theta_d  + \cos \theta_e \cos \theta_d \;.
\end{equation}
For the error function we use the convention
\begin{equation}
\erf(z) = \frac{2}{\sqrt{\pi}} \int_0^z \text{e}^{-t^2} \diff t
\end{equation}
The normalisation factor for the energy smearing is rather straight-forward
to derive and we will not comment on it any further.
For the angular part this is less trivial.
Let us first note that it is rather straight-forward to
do the relevant integral for $\hat{q}_d$ just pointing in the $z$-direction. Then
$\hat{q}_e \cdot \hat{q}_d =  \cos \theta_e$ and
\begin{align}
\int \diff \Omega_e \, f_\theta &= 2 \pi \mathcal{N}_\theta \int_{0}^{\pi} \diff \theta_e \, \sin \theta_e  \exp\left( - \frac{ \theta_e^2 }{2 \, \sigma_\theta^2}\right) \nonumber\\
&= \mathcal{N}_\theta\sqrt{\frac{\pi^3 }{2}} \exp\left(-\frac{\sigma_\theta^2}{2}\right)
\frac{\sigma_\theta}{\ci}
\left(
2 \erf\left(\frac{\ci \sigma_\theta}{\sqrt{2}} \right)
-\erf\left(\frac{\ci \sigma_\theta^2 + \pi }{\sqrt{2}\sigma_\theta}\right)
-\erf\left(\frac{\ci \sigma_\theta^2 - \pi }{\sqrt{2} \sigma_\theta}\right)
\right) \nonumber\\
&\stackrel{!}{=} 1\;,
\end{align}
from which we can read off the normalisation factor easily.
But the general case is related to this one by a coordinate
transformation and we always integrate over the whole
solid angle. Therefore the normalisation constant does not
depend on $\theta_d$ and $\phi_d$.
That is different to the energy case since the normalisation
there depends on how far
$E_d$ is from the physical threshold.

\end{appendix}

\end{document}